\documentclass[leqno,11pt]{article}  

\usepackage{lineno}
\usepackage{lmodern}

\usepackage{algorithm}
\usepackage[noend]{algorithmic}
\usepackage{algorithmicext}
\usepackage{ifthen}
\newconstruct{\PROC}{\textbf{procedure}}{}{\ENDPROC}{\textbf{end procedure}}
\newconstruct{\FUNC}{\textbf{function}}{}{\ENDFUNC}{\textbf{end function}}
\usepackage{amsmath,amssymb,amsthm}
\usepackage{graphicx}
\usepackage{color}
\usepackage{mathrsfs }
\usepackage[normalem]{ulem}
\usepackage{paralist}
\newcommand{\msg}[1]{\langle#1\rangle}
\renewcommand{\epsilon}{\varepsilon}

\usepackage{fullpage}

\DeclareMathAlphabet{\mathsc}{OT1}{cmr}{m}{sc}

\newtheorem{theorem}{Theorem}[section]
\newtheorem{lemma}[theorem]{Lemma}
\newtheorem{corollary}[theorem]{Corollary}

\newtheorem{property}{Property}

\newtheorem{observation}{Observation}

\renewcommand{\geq}{\geqslant}
\renewcommand{\ge}{\geqslant}
\renewcommand{\leq}{\leqslant}
\renewcommand{\le}{\leqslant}
\newcommand{\set}[1]{\{#1\}}
\newcommand{\la}{\leftarrow}

\newcommand{\eps}{\varepsilon}

\newcommand{\dd}{{\mathsc{DD}}}
\newcommand{\infl}{{\mathsc{Infl}}}

\newcommand{\val}{{\mathsc{val}}}
\newcommand{\lo}{{\mathsc{Low}}}
\newcommand{\hi}{{\mathsc{High}}}
\renewcommand{\mid}{{\mathsc{Mid}}}

\newcommand{\A}{{\mathcal{A}}}

\renewcommand{\L}{{\mathcal{L}}}

\newcommand{\V}{{\mathcal{V}}}
\newcommand{\R}{{\mathcal{R}}}
\newcommand{\sa}{{\sc Stable Agreement}}

\newcommand{\bc}{{\sc Binary Consensus}}

\newboolean{short}
\setboolean{short}{false}

\newcommand{\shortOnly}[1]{\ifthenelse{\boolean{short}}{#1}{}}
\newcommand{\onlyShort}[1]{\ifthenelse{\boolean{short}}{#1}{}}
\newcommand{\longOnly}[1]{\ifthenelse{\boolean{short}}{}{#1}}
\newcommand{\onlyLong}[1]{\ifthenelse{\boolean{short}}{}{#1}}

\newcommand*\patchAmsMathEnvironmentForLineno[1]{%
  \expandafter\let\csname old#1\expandafter\endcsname\csname #1\endcsname
  \expandafter\let\csname oldend#1\expandafter\endcsname\csname end#1\endcsname
  \renewenvironment{#1}%
     {\linenomath\csname old#1\endcsname}%
     {\csname oldend#1\endcsname\endlinenomath}}%
\newcommand*\patchBothAmsMathEnvironmentsForLineno[1]{%
  \patchAmsMathEnvironmentForLineno{#1}%
  \patchAmsMathEnvironmentForLineno{#1*}}%
\AtBeginDocument{%
\patchBothAmsMathEnvironmentsForLineno{equation}%
\patchBothAmsMathEnvironmentsForLineno{align}%
\patchBothAmsMathEnvironmentsForLineno{flalign}%
\patchBothAmsMathEnvironmentsForLineno{alignat}%
\patchBothAmsMathEnvironmentsForLineno{gather}%
\patchBothAmsMathEnvironmentsForLineno{multline}%
}

\begin{document}
\title{Distributed Agreement in Dynamic Peer-to-Peer Networks\footnote{A preliminary version of this paper appeared in the Proceedings of the  ACM/SIAM Symposium on Discrete Algorithms (SODA), 2012, 551-569.}}
\author{John Augustine\thanks{Department of Computer Science and Engineering, Indian Institute of Technology Madras, Chennai, India.  \hbox{E-mail}:~{\tt augustine@cse.iitm.ac.in}. Work done while at the 
Division of Mathematical Sciences, Nanyang Technological University, Singapore 637371.}  \and Gopal Pandurangan \thanks{Division of Mathematical
Sciences, Nanyang Technological University, Singapore 637371 and Department of Computer Science, Brown University, Box 1910, Providence, RI 02912, USA.  \hbox{E-mail}:~{\tt gopalpandurangan@gmail.com}. Work supported in part by the following grants: Nanyang Technological University grant M58110000, Singapore Ministry of Education (MOE) Academic Research Fund (AcRF) Tier 2 grant MOE2010-T2-2-082,
US NSF grant CCF-1023166, and a grant from the US-Israel Binational Science
Foundation (BSF).} \and Peter Robinson \thanks{Department of Computer Science, National University of Singapore. \hbox{E-mail}:~{\tt robinson@comp.nus.edu.sg}}
\and Eli Upfal\thanks{Department of Computer Science, Brown University, Box 1910,
Providence, RI 02912, USA. \hbox{E-mail}:~{\tt eli@cs.brown.edu}}}

\date{}

\maketitle

\begin{abstract}
Motivated by the need for robust and fast distributed computation in highly dynamic Peer-to-Peer (P2P)
networks, we study algorithms for  the fundamental  distributed agreement problem.   P2P
networks  are  highly dynamic networks  that experience heavy
node {\em  churn} (i.e., nodes  join and leave the network continuously over time). Our
goal  is to design fast  algorithms (running in a small number of rounds) that
guarantee, despite high node churn rate,  that almost all nodes   reach a stable
agreement.  Our main contributions are randomized  distributed algorithms that
guarantee   {\em stable almost-everywhere agreement} with high probability even under
high adversarial churn in a polylogarithmic number of rounds. In particular, we
present the following results:
\begin{enumerate}
\item An $O(\log^2 n)$-round  ($n$ is the stable network size) randomized
  algorithm that achieves almost-everywhere agreement with high probability
  under up to {\em linear} churn  {\em per round} (i.e., $\epsilon n$, for some small
  constant $\epsilon > 0$), assuming that the churn is controlled by an
  oblivious adversary (that has  complete knowledge and control of what nodes join
  and leave and  at what time  and has unlimited computational power, but is
  oblivious to the random choices made by the algorithm). Our algorithm requires 
only polylogarithmic  in $n$  bits to be processed and sent (per round) by each node.
\item An  $O(\log m\log^3 n)$-round randomized algorithm that achieves
  almost-everywhere agreement with high probability under up to $\epsilon
  \sqrt{n}$ churn  per round (for some small $\epsilon > 0$), where $m$ is the size of the input value domain, that works even under an adaptive adversary
  (that also knows the past random choices made by the algorithm). This algorithm requires up to polynomial in $n$ bits (and up to $O(\log m)$ bits)
to be processed and sent (per round) by each node. 
\end{enumerate}

Our algorithms are the first-known, fully-distributed,  agreement algorithms that work under highly dynamic settings (i.e., high churn rates per step). Furthermore, they are  localized (i.e., do not require any global topological knowledge), simple, and easy to implement. 
 These algorithms can serve as building blocks for implementing other non-trivial
distributed computing tasks in dynamic P2P networks. 
\end{abstract}

\section{Introduction}

Peer-to-peer (P2P) computing is emerging as one of the key networking technologies in recent years with many application systems, e.g., Skype, BitTorrent, Cloudmark etc. However, many of these  systems are not truly P2P, as they are not
fully decentralized --- they typically use hybrid P2P along with centralized intervention.
For example, Cloudmark  \cite{Cloudmark} is a large spam detection system used by millions of people that operates by maintaining
a hybrid P2P network; it uses a central authority to regulate and charge users for participation in the network.
A key reason for the lack of fully-distributed P2P systems is  the difficulty in designing highly robust algorithms for large-scale dynamic P2P networks. Indeed, P2P networks  are  highly dynamic  networks  characterized by  high degree of node {\em  churn}  --- i.e., nodes continuously join and leave the network. Connections (edges) may be added or deleted at any time and thus the topology changes very dynamically. In fact,  
measurement studies of real-world P2P networks~\cite{FPJKA07,SGG02,SW02, SR06} show that the churn rate is quite high:
nearly 50\% of peers in real-world networks can be replaced within an hour. (However,  despite a large churn rate, these studies  also show that the
total number of peers in the network is relatively {\em stable}.)  We note that peer-to-peer algorithms have been proposed
 for a wide variety
of  computationally challenging tasks such as collaborative
filtering~\cite{Canny02}, spam detection~\cite{Cloudmark},
data mining~\cite{DBGWK06}, worm detection and
suppression~\cite{MS05,VAS04}, and privacy protection of archived
data~\cite{GKLL09}. However, all algorithms proposed for these
problems have no theoretical guarantees of being able to
work in a network with a dynamically changing topology and a linear churn rate per round. This is  a major bottleneck in implementation and
wide-spread use of these algorithms.

In this paper, we take a step towards  designing robust  algorithms for large-scale dynamic
peer-to-peer networks. In particular, we  study the fundamental  distributed agreement problem in  P2P networks (the formal problem statement and model is given in Section \ref{sec:model}). An efficient solution to the agreement problem can be used as a building block for robust and efficient solutions to other problems as mentioned above. 
However, the distributed agreement problem in P2P networks is challenging since the goal
is to guarantee {\em almost-everywhere} agreement, i.e., almost all nodes
\footnote{In sparse, bounded-degree networks,
an adversary can always isolate some number of non-faulty nodes, hence almost-everywhere is
the best one can hope for in such networks \cite{DPPU88}.} 
should reach consensus,  even under high churn rate. The churn rate
can be as much as linear  {\em per time step (round)},  i.e., up to a constant fraction of the stable network size can be replaced per time step.  
Indeed, until recently, almost all the work 
 known in the literature (see e.g., \cite{DPPU88,KKKSS10, KS10,KSS06, Upfal94})  have addressed the almost-everywhere agreement problem only in
 static (bounded-degree) networks and these approaches do not work for dynamic networks with changing topology. 
Such approaches fail in dynamic networks where both nodes {\em and} edges can change by a large amount
in {\em every} round.
   For example, the work of Upfal \cite{Upfal94} showed how
 one can achieve almost-everywhere agreement under up to a {\em linear} number --- up to $\epsilon n$, for a sufficiently small $\epsilon > 0$ ---  of Byzantine faults in a bounded-degree expander network ($n$ is the network size).  The algorithm required $O(\log n)$ rounds and polynomial (in $n$) number of messages; however,  the local computation required by
each processor is exponential.  Furthermore, the algorithm  requires  knowledge
of the global topology, since at the start, nodes need to have this information
``hardcoded''.
 The work of King et al. \cite{KSSV06} is important in the context of P2P networks, as it was the first to study scalable (polylogarithmic communication and number of rounds) algorithms for distributed agreement (and leader election)
that are tolerant to Byzantine faults. However, as pointed out by the authors, their algorithm works only for static networks; similar to Upfal's algorithm, the nodes require 
hardcoded information on the network topology to begin with  and thus the
algorithm does not work when the topology changes.  In fact, this work (\cite{KSSV06})
raises the open question of whether one can design agreement  protocols that
can work in highly dynamic networks with a large churn rate.

\subsection{Our Main Results}
 Our first contribution is a
rigorous theoretical framework for the design and analysis of algorithms for
highly dynamic distributed systems with churn. We briefly describe the key ingredients of our model here. (Our model is described in detail in  Section
\ref{sec:model}.) Essentially, we model a P2P network as a bounded-degree expander graph   whose topology --- both nodes and edges ---  can change arbitrarily from round to round and is controlled by an adversary. However, we assume
that the total number of nodes in the network is stable.  The number of node
changes {\em per round} is called the {\em churn rate} or {\em churn limit}. We
consider a churn rate of up to some $\epsilon n$,
where $n$ is the stable network size. Note that our model is quite general in the
sense that we only assume that the topology is an expander at every step; no other special
properties are assumed. Indeed,  expanders have been used extensively to
model dynamic P2P networks\footnote{Expander graphs
have been used extensively as candidates to solve the agreement and related problems
in bounded degree graphs even in static settings (e.g., see \cite{DPPU88,KKKSS10, KS10,KSS06, Upfal94}).  Here we show that similar expansion properties are beneficial in the more challenging setting of dynamic networks.}   in which the expander property is preserved under insertions and
deletions of nodes (e.g., \cite{LS03,PRU01}). 
 Since we do not make assumptions on how the topology is preserved,
 our model is applicable to all such expander-based networks.  (We note that  various prior work on  dynamic network models make similar assumptions on preservation of  topological  properties  --- such as connectivity, expansion etc. ---
 at every step  under dynamic {\em edge} insertions/deletions --- cf. Section \ref{sec:related}.  The issue
 of how such properties are preserved are abstracted away from the model, which allows one to focus on
 the dynamism. Indeed, this abstraction has been a feature of most dynamic
 models e.g., see the survey  of \cite{santoro}.) 
 
We study stable, almost-everywhere,  agreement in our model. By ``almost-everywhere'',
we mean that almost all nodes, except possibly $ \beta c(n)$ nodes (where $c(n)$ is the order of the churn and $\beta > 0$ is a suitably small constant --- cf. Section~\ref{sec:model}) should reach agreement on a common value.
(This agreed value must be the input value of some node.) By ``stable'' we mean that the agreed value is preserved subsequently after the agreement is reached.
 
Our main contribution is the design and analysis of randomized  distributed
algorithms that guarantee   stable almost-everywhere agreement with high
probability (i.e., with probability $1 - 1/n^{\gamma}$, for an arbitrary fixed constant $\gamma\ge 1$) even under high
adversarial churn in a polylogarithmic number of
rounds.  Our algorithms also  guarantee stability once agreement has been
reached.  In particular, we present the following
results (the precise theorem statements are given in the respective sections
below):
\begin{enumerate}
\item (cf.\ Section \ref{sec:oblivious}) An $O(\log^2 n)$-round  ($n$ is the stable
network size) randomized algorithm that achieves almost-everywhere
agreement with high probability under up to {\em linear} churn {\em per round}
(i.e., $\epsilon n$, for some small constant $\epsilon > 0$), assuming
that the churn is controlled by an oblivious adversary (that has complete
knowledge of what nodes join and leave and  at what time, but is oblivious
to the random choices made by the algorithm). Our algorithm requires 
only polylogarithmic  in $n$  bits to be processed and sent (per round) by each node.

\item (cf.\ Section \ref{sec:adaptive}) An  $O(\log m \log^3 n)$-round
randomized algorithm that achieves almost-everywhere agreement with high
probability under up to $\epsilon \sqrt{n}$ churn {\em per round}, for some
small $\epsilon > 0$, that works even under an adaptive adversary (that also
knows the past random choices made by the algorithm). Here $m$ refers
to the size of the domain of input values. This algorithm requires up to polynomial in $n$ bits (and up to $O(\log m)$ bits)
to be processed and sent (per round) by each node. 

\item (cf.\ Section \ref{sec:impossibility}) We also show that no
deterministic algorithm can guarantee almost-everywhere agreement
(regardless of the number of rounds), even under constant churn rate. 
\end{enumerate}

To the best of our knowledge, our algorithms are the first-known,
fully-distributed,  agreement algorithms that work under highly dynamic
settings. Our algorithms are  localized (do not require any global topological
knowledge), simple, and easy to implement. 
 These algorithms can serve as building blocks for implementing other non-trivial
distributed computing tasks in P2P networks. 

\onlyShort{
Due to lack of space, the full proofs are in the full paper \cite{APRU11}.
}

\subsection{Technical Contributions}
The main technical challenge that we have to overcome is designing and analyzing
distributed algorithms in networks where both nodes and edges can change by a large amount.
Indeed, when the churn rate is linear, i.e., say $\epsilon n$ per round, in constant ($1/\epsilon$) number of rounds the entire network can be renewed!
 
We derive techniques for information spreading  (cf.\ Section
\ref{sec:techniques}) for doing non-trivial distributed computation in such networks. The first technique that we use
is flooding.   We show that in an expander-based
P2P network even under linear churn rate, it is possible to spread
information by flooding if  sufficiently many (a $\beta$-fraction of the
order of the churn)
nodes initiate the information spreading (cf.~Lemma~\ref{lem:beta}). In other words,  even an adaptive adversary 
cannot ``suppress'' more than  a small fraction
of the values. The precise statements
and proofs are in Section \ref{sec:techniques}. 

To analyze these flooding techniques we introduce the dynamic
distance, which describes the effective distance between two nodes with respect
to the causal influence. We define the notions of influence sets and dynamic
distance (or flooding time) in dynamic networks with node churn.  (Similar
notions have been defined for dynamic graphs with
a fixed set of nodes, e.g., \cite{kuhn-survey,BCF09}).  In (connected) networks
where the nodes are fixed, the effective diameter (e.g., \cite{kuhn-survey}) is
always finite. In the highly dynamic setting considered here, however, the
effective distance between two nodes might be infinite, thus we need a more
refined definition for influence set and dynamic distance.

The second technique that we use is ``support estimation'' (cf.\ Section
\ref{sec:support}). Support estimation is a randomized technique that allows us to estimate the aggregate count (or sum)
of values of all or a subset of nodes in the network. Support estimation is done in conjunction with flooding
and uses properties of the exponential distribution (similar to \cite{Cohen97, AoyamaS08}).
Support estimation allows us to estimate the aggregate value quite precisely
with high probability even under linear churn. But this works only for an
oblivious adversary; to get similar results for the adaptive case, we need to
increase the amount of bits that can be processed and sent by a node in every
round. 

Apart from support estimation, we also use our flooding techniques in the
agreement algorithm for the oblivious case (cf.\ Algorithm~\ref{alg:bc}) to sway
the decision one way or the other. \onlyLong{For the adaptive case (cf.\
Algorithm~\ref{alg:adaptive}), we use the variance
property of a certain probability distribution to achieve the same effect with
constant probability.}

\subsection{Other Related Work}
\label{sec:related}
\subsubsection{Distributed Agreement}
The distributed agreement (or consensus) problem is important in a wide range
of applications, such as database management, fault-tolerant analysis of
aggregate data, and coordinated control of multiple agents or peers. 
There is a long line of research on various versions of the 
problem with many important results (see e.g., \cite{AW04,Lyn96}
and the references therein). 
The relaxation of achieving agreement ``almost everywhere'' was introduced
by \cite{DPPU88} in the context of fault-tolerance in networks of bounded degree where all
but $O(t)$ nodes achieve agreement despite $t=O(\frac{n}{\log n})$ faults.
This result was improved by \cite{Upfal94}, which showed how to guarantee
almost everywhere agreement in the presence of a linear fraction of faulty
nodes. Both the work of \cite{DPPU88, Upfal94} crucially use expander graphs to show their results. We also refer to the related results of Berman and Garay on the
butterfly network~\cite{BG93}. 

\subsubsection{Byzantine Agreement}
We note that Byzantine adversaries are quite different from the adversaries considered in this paper.  A Byzantine
adversary can have nodes behaving arbitrarily, but no new nodes  are added (i.e., no churn),
whereas in our case (an external) adversary  controls the churn and topology of
the network but {\em not} the behavior of the nodes. Despite this difference it is worthwhile to mention that
there has been significant work in designing peer-to-peer networks that
are provably robust to a large number of Byzantine
faults~\cite{FS02,HK03,NW03,Scheideler05}. These focus only on robustly enabling storage and retrieval of data items. 
The problem of achieving almost-everywhere agreement among nodes in P2P networks (modeled as an expander graph) is
considered by King et al.\ in \cite{KSSV06} in the context of the leader
election problem; essentially, \cite{KSSV06} is a sparse (expander) network
implementation of the full information protocol of \cite{KSS06}.  
More specifically, \cite{KSSV06} assumes that the
adversary corrupts a constant fraction $b < 1/3$ of the
processes that are under its control throughout the run of the algorithm.
The protocol of \cite{KSSV06} guarantees that with constant probability an
uncorrupted leader will be elected and that a $1-O(\frac{1}{\log n})$
fraction of the uncorrupted processes know this leader. Again, we note that the
failure assumption of \cite{KSSV06} is quite different from the one we use: Even though we do not assume corrupted nodes, the adversary is free
to subject different nodes to churn in every round. Also note that
the algorithm of \cite{KSSV06} does not work for dynamic networks. 

Other works on handling Byzantine nodes in the context of P2P networks include \cite{Scheideler05,awerbuch:group,fiat:dynamically,fiat:making,awerbuch:random,castro:secure,young:practical}.

In \cite{podc13}, we have developed an almost-everywhere agreement algorithm that tolerates up to $\tilde O(\sqrt{n})$ churn and $\tilde O(\sqrt{n})$ churn per round, in a dynamic network model.

\subsubsection{Dynamic Networks}
Dynamic networks have been studied extensively over the past three
decades.  Some of the early studies focused on dynamics that arise out
of faults, i.e., when edges or nodes fail.  A number of fault models,
varying according to extent and nature (e.g., probabilistic
vs.\ worst-case) and the resulting dynamic networks have been analyzed
(e.g., see~\cite{AW04,Lyn96}).  There have
been several studies on models that constrain the rate at which
changes occur, or assume that the network eventually stabilizes (e.g.,
see~\cite{afek+ag:dynamic,dolev:stabilize,gafni+b:link-reversal}). 
Some of the early work on general dynamic networks
include~\cite{afek+gr:slide,awerbuch+pps:dynamic} which introduce
general building blocks for communication protocols on dynamic
networks.  Another notable work is the local
balancing approach of~\cite{awerbuch+l:flow} for solving routing and
multicommodity flow problems on dynamic networks.
Most of these papers develop algorithms that will work under the assumption that 
the network will eventually stabilize and stop changing.

Modeling general dynamic networks has gained renewed attention with
the recent advent of heterogeneous networks composed out of ad hoc,
and mobile devices. To address highly unpredictable network dynamics, stronger adversarial
models have been studied
by~\cite{avin+kl:dynamic,disc12,OW05,KOM11} and others;
see the recent survey of \cite{santoro} and the references therein. The works of  \cite{KOM11, avin+kl:dynamic, disc12} study a model
in which the communication graph can change completely from one round
to another, with the only constraint being that the network is
{\em connected at each round} (\cite{KOM11} and \cite{disc12} also consider a stronger model where the constraint is that
the network should be an expander or should have some specific expansion  in each round).
The model has also been applied to agreement problems in dynamic networks;
various versions of coordinated consensus (where all nodes must agree) have been
considered  in \cite{KOM11}. 
  The recent work of \cite{clementi-podc12},
studies the flooding time of {\em Markovian} evolving dynamic graphs,
a special class of evolving graphs.
 
We note that the model of~\cite{kuhn+lo:dynamic}
allows only edge changes from round to round while the nodes remain
fixed. In this work, we introduce a dynamic
network model  where both nodes and edges
can change by a large amount (up to a linear fraction of the network
size).  
Therefore, the framework we introduce in Section~\ref{sec:model} is more general
than the model of \cite{kuhn+lo:dynamic}, as it is additionally applicable to dynamic
settings with node churn.
The same is true for the notions of dynamic distance and influence
set that we introduce in Section~\ref{sec:dynamic}, since in our model the
dynamic distance is not necessarily finite. In fact, according to
\cite{kuhn-survey}, coping with churn is one of the important open problems in the
context of dynamic networks.  Our paper takes a step in this direction.

An important aspect of our algorithms  is that they will work and terminate correctly even when the
  network keeps continually changing.  We note that there has been
considerable prior work in dynamic P2P networks (see \cite{PRU01} and the references therein) but these do not assume that
the network keeps continually changing over time.

Due to the mobility of nodes, mobile ad-hoc networks can also be considered as
dynamic networks. The focus of \cite{OW05} are the minimal requirements
that are necessary to correctly perform flooding and routing in highly dynamic
networks where edges can change but the set of nodes remains the same. In the
context of agreement problems, electing a leader among mobile nodes that may
join or leave the network at any time is the focus of \cite{CRW11}. To make
leader election solvable in this model, Chung et al.\ introduce the notion of
$D$-connectedness, which ensures information propagation among all nodes that
remain long enough in the network. Note that, in contrast to our model, this
assumption prohibits the adversary from permanently isolating parts of the
network. The recent work of~\cite{haeupler+k:dynamic} presents
information spreading algorithms on dynamic networks based on network
coding~\cite{ahlswede+cly:coding}. 

\subsubsection{Fault-Tolerance}

In most work on fault-tolerant agreement problems 
the adversary a priori commits to a fixed set of faulty nodes.
In contrast, \cite{DGMSS11} considers an adversary that can corrupt the
state of some (possibly changing) set of $O(\sqrt{n})$ nodes in every round.
The median rule of \cite{DGMSS11} provides an elegant way to ensure that
most nodes stabilize on a common output value within $O(\log n)$ rounds,
assuming a complete communication graph. The median rule, however, only
guarantees that this agreement lasts for some polynomial number of rounds,
whereas we are able to retain agreement ad infinitum.

Expander graphs and spectral properties have already been applied
extensively to improve the network design and fault-tolerance in
distributed computing (cf.\ \cite{Upfal94,DPPU88,BBCES2006}).
Law and Siu\ \cite{LS03} provide a distributed algorithm for maintaining an expander in
the presence of churn with high probability by using Hamiltonian cycles.
In \cite{PT11} it is shown how to maintain the expansion property of a network
in the self-healing model where the adversary can delete/insert a new node in
every step.
In the same model, \cite{IPDPS14} present a protocol that maintains constant node degrees and constant expansion (both with probability $1$) against an adaptive adversary, while requiring only logarithmic (in the network size) messages, time, and topology changes per deletion/insertion.
In \cite{skipexp}, it is shown that a SKIP graph (cf.\ \cite{skip}) contains a constant degree expander as a subgraph with high probability.
Moreover, it requires only constant overhead for a node to identify its incident edges that are part of this expander.
Later on, \cite{skipplus} presented a self-stabilizing algorithm that converges from any weakly connected graph to a SKIP graph in time polylogarithmic in the network size, which yields a protocol that constructs an expander with high probability.
In  \cite{hyperring} the authors introduce the hyperring, which is a search data structure supporting insertions and deletions, while being able to handle concurrent requests with low congestion and dilation, while guaranteeing $O(1/\log n)$ expansion and $O(\log n)$ node degree.
The $k$-Flipper algorithm of \cite{mahlmann} transforms any undirected graph into an expander (with high probability) by iteratively performing flips on the end-vertices of paths of length $k+2$.
Based on this protocol, the authors describe how to design a protocol that supports deletions and insertions of nodes.
Note that, however, the expansion in \cite{mahlmann} is only guaranteed with high probability however, assuming that the node degree is $\Omega(\log n)$.

Information spreading in distributed networks is the focus of
\cite{CH:PODC10} where it is shown that this problem requires $O(\log n)$
rounds in graphs with a certain conductance in the push/pull model 
where a node can communicate with a
randomly chosen neighbor in every round. 

Aspnes et al.\ \cite{ARS07} consider information spreading via expander
graphs against an adversary, which is related to the flooding
techniques we derive in Section~\ref{sec:techniques}. More specifically,
in \cite{ARS07} there are two opposing parties ``the alert'' and ``the
worm'' (controlled by the adversary) that both try to gain control of the
network. In every round each alerted node can alert a constant number of
its neighbors, whereas each of the worm nodes can infect a constant
number of non-alerted nodes in the network. In \cite{ARS07}, Aspnes et al.\ show that there is a simple strategy to prevent all
but a small fraction of nodes from becoming infected and, in case that the
network has poor expansion, the worm will infect almost all nodes.

The work of \cite{BBCES2006} shows that, given a network that is initially
an expander and assuming some linear fraction of faults, the remaining
network will still contain a large component with good expansion.  These
results are not directly applicable to dynamic networks with large amount
of churn like the ones we are considering, as the topology might be
changing and linear churn per round essentially corresponds to
$O(n\log n)$ total churn after $\Theta(\log n)$ rounds---the minimum amount
of time necessary to solve any non-trivial task in our model.

In the context of maintaining properties in P2P networks, Kuhn et al.\
consider in \cite{KSW10} that up to $O(\log n)$ nodes can crash or join per constant number of time
steps. Despite this amount of churn, it is shown in \cite{KSW10} how to maintain a low peer
degree and bounded network diameter in P2P systems by using the hypercube and
pancake topologies.  Scheideler and Schmid show in \cite{SS09} how to maintain a
distributed heap that allows join and leave operations and, in addition, is
resistent to Sybil attacks. A robust distributed implementation of a distributed
hash table (DHT) in a P2P network is given by \cite{AS09}, which can withstand
two important kind of attacks: adaptive join-leave attacks and adaptive
insert/lookup attacks by up to $\eps n$ adverserial peers. 
Note that, however, that collisions are likely to occur once the number of attacks becomes $\Omega(\sqrt{n})$.

\section{Model and Problem Statement} \label{sec:model}

We are interested in establishing  stable agreement in a
dynamic peer-to-peer network in which the nodes and the edges change over time.
The computation is structured into synchronous rounds, i.e., we assume that
nodes run at the same processing speed and any message that is sent by some node
$u$ to its (current) neighbors in some round $r\ge 1$ will be received by the end of $r$.
To ensure scalability, we restrict the number of bits sent per round by each
node to be polylogarithmic in the size of the input value domain (cf.\
Section~\ref{sec:sa}). For dealing with the
much more powerful adaptive adversary, we relax this requirement in
Sections~\ref{sec:supportAdaptive} and \ref{sec:adaptive}.
We model dynamism in the network as a family of undirected graphs $(G^r)_{r\ge
0}$. At the beginning of each round $r$ we start with the network topology
$G^{r-1}$. Then, the adversary gets to change the network from $G^{r-1}$ to
$G^r$ (in accordance to rules outlined below). 
As is typical, an
edge $(u,v) \in E^r$ indicates that $u$ and $v$ can communicate in round $r$ by
 passing messages. 
For the sake of
readability, we use $V^{[r,r+t]}$ as a shorthand for
$\bigcap_{i=r}^{r+t}V^i\text{.}$ 
Each node $u$ has a unique identifier and is {\em churned in} at some round
$r_i$ and {\em churned out} at some $r_o > r_i$. More precisely, for each node
$u$, there is a maximal range $[r_i, r_o -1]$ such that $u \in V^{[r_i, r_o-1]}$ and
for every $r \notin [r_i, r_o-1]$, $u \notin V^r$. Any information about the
network at large is only learned through
the messages that $u$ receives. It has no a priori knowledge about who its neighbors will be in
the future. Neither does $u$ know when (or whether) it will be churned out. 
Note that we do not assume that nodes have access to perfect clocks\onlyLong{,
but we show (cf.  Section~\ref{subsec:maintain}) how the nodes can synchronize
their clocks}. 

We make the following assumptions about the kind of changes that our dynamic
network can encounter:

\begin{description}
\item[Stable Network Size:] For all $r$, $|V^r| = n$, where $n$ is a suitably
  large positive integer.  This assumption simplifies our analysis. Our algorithms will work correctly as long as the number of nodes is reasonably stable (say, between $n - \kappa n$ and $n+\kappa n$ for some suitably small constant  $\kappa$). Also, we assume that $n$ (or a constant factor estimate of $n$) is common knowledge among the nodes in
  the network\footnote{This assumption is important; estimating $n$ accurately in our model is an interesting problem in itself.}.
\item[Churn:]  For each $r>1$, 
  $$|V^r \setminus V^{r-1}| = |V^{r-1} \setminus V^{r}| \le \L = \eps c(n)
  \text{,}$$
where $\L$ is the {\em churn limit}, which is some fixed $\eps>0$ fraction of the
\emph{order of the churn} $c(n)$; the equality in the above equation
ensures that the network size remains stable.  Our work is aimed at high levels
of churn up to a churn limit $\L$ that is linear in $n$, i.e., $c(n) = n$. 
\item[Bounded Degree Expanders:] The sequence of graphs $(G^r)_{r\geq 0}$ is an
  expander family with a vertex expansion of at least $\alpha$, which is a fixed positive constant.\footnote{Note that the value of $\alpha$ determines $\eps$, i.e.\ the fraction of churn that we can tolerate. In particular, to tolerate
 linear amount of churn, we require constant expansion.  In principle, our results can potentially be extended to graphs with weaker expansion guarantees as well; however the amount of churn that can be tolerated will
 be reduced.}
  In
  other words, the adversary must ensure that for every $G^r$ and  every $S \subset V^r$ such that $|S| \le n/2$,
  the number of nodes in $V^r \setminus S$ with a neighbor in $S$ is at least
  $\alpha |S|$.   
  Note that we do not explicitly consider the costs (communication and computation) of maintaining an expander under churn.
  Instead, we assume that the duration of each time step in our model are normalized to be large enough to encompass an expander maintenance protocol such as \cite{LS03,IPDPS14}.
  
\end{description}

A run of a distributed algorithm consists of an infinite number of
rounds. 
We assume that the following events occur (in order) in every round $r$:
\begin{enumerate}
    \item  A set of at most $\L$ nodes are churned in and another set of $\L$
      nodes are churned out. The edges of $G^{r-1}$ may be changed as well, but
      $G^r$ has to have a vertex expansion of at least $\alpha$. These changes
      are under the control of the adversary.
		\item  The nodes broadcast messages to their (current) neighbors. 
    \item  Nodes receive messages broadcast by their neighbors.
    \item  Nodes perform computation that can change their state and determine
      which messages to send in round $r+1$.
		\end{enumerate}

\subsection*{Bounds on Parameters}
Recall that the churn limit $\L = \eps c(n)$, where $\eps > 0$ is a constant and $c(n)$ is the churn order. When $c(n) = n$,  $\eps$ is the fraction of the nodes churned out/in and therefore we require $\eps$  to be less than 1 and must adhere to Equation~\eqref{eq:beta}.
Moreover, we require the bound $\beta < \frac{1}{12}$ regarding the right hand side of \eqref{eq:beta}.
However, when $c(n) \in o(n)$,  $\eps$ can exceed 1. In the remainder of this paper, we consider $\beta$ to be a  small
constant independent of $n$, such that

\begin{equation} \label{eq:beta}
  \frac{\eps (1 + \alpha)}{\alpha} < \beta.
\end{equation}
It will become apparent in Section~\ref{sec:techniques} that \eqref{eq:beta} presents a sufficient condition for preventing the adversary from containing the information propagated by a set of $\beta c(n)$ nodes.

and that the \emph{churn expansion ratio}
$\frac{\eps(1 + \alpha)}{\alpha}$ presents a sufficient condition for information
propagation in our model (cf.\ Lemma~\ref{lem:beta}).
Finally, we assume that $n$ is suitably large (cf. Equations~\ref{eq:betaUpperOblivous}~and~\ref{eq:numberOfNodesAdaptive}). 

\subsection{Stable Agreement} \label{sec:sa}
We now define the {\sc Almost Everywhere} \sa\ problem (or just the \sa\ problem for brevity). Each node $v \in V^0$ has an associated input
value from some value domain of size $m$; subsequent new nodes come with value $\bot$. 
Let $\V$ be the set of all input values associated with nodes in $V^0$ at the start of round
1. Every node $u$ is equipped with a special decision variable $decision_u$ (initialized to $\bot$) that can be written at
most once. We say that a node $u$ \emph{decides on $\val$} when $u$ assigns $\val$ to its $decision_u$. Note that this decision is
irrevocable, i.e., every node can decide at most once in a run of an algorithm. As long as $decision_u = \bot$, we say that $u$ is {\em undecided}.
\sa\ requires that a large fraction of the nodes come to a stable
agreement on one of the values in $\V$. More precisely, \emph{an algorithm solves
\sa\ in $R$ rounds}, if it exhibits the following characteristics in every run,
for any fixed $\beta$ adhering to \eqref{eq:beta}.
\begin{description}
\item[Validity:] If, in some round $r$, node $u \in V^r$ decides on a value
  $\val$, then $\val \in \V$. 
\item[Almost Everywhere Agreement:] We say that \emph{the network has
  reached strong almost everywhere agreement by round $R$}, if at least $n-\beta
  c(n)$ nodes in $V^R$ have decided on the same value $\val^* \in \V$ and every
  other node remains undecided, i.e., its decision value is $\bot$. In particular, no
  node ever decides on a value $\val'\in \V$ in the same run, for $\val'\ne\val^*$.
\item[Stability:] Let $R$ be the earliest round where nodes have reached almost
  everywhere agreement on value $\val^*$. We say that an algorithm \emph{reaches stability by round $R$} if, at every
  round $r \geq R$, at least $n-\beta c(n)$ nodes in $V^r$ have decided on
  $\val^*$.
\end{description}
We also consider a weaker variant of the above problem that we call 
{\sc Almost Everywhere} \bc\ (or simply, \bc) where
the input values in $\V$ are restricted to $\set{0,1}$.

We consider two types of adversaries for our randomized algorithms. An
\emph{oblivious} adversary  must commit in advance to the entire sequence of
graphs $(G^r)_{r\ge 0}$. In other
words, an oblivious adversary must commit independently of the random choices
made by the algorithm. 
We also consider the more powerful
  \emph{adaptive} adversary that can observe the entire state of the network in
  every round $r$ (including all the random choices made until  round $r-1$), and then chooses the nodes to be churned out/in and how to
  change the topology of $G^{r+1}$.

For the sake of readability, we treat $\log n$ as an integer and omit the
necessary ceiling or floor operations if their application is clear from the
context.

\section{Techniques for Information Spreading} \label{sec:techniques}
In this section, we first derive and analyze techniques to spread information in the network despite churn.
First, we show that the adversary is unable to prevent a sufficiently large set of nodes (of size at least $\beta c(n)$) to propagate their information to almost all other nodes (cf.\ Lemma~\ref{lem:beta}). 
Building on this result, we analyze the capability of individual nodes to spread their information.
We show in Lemma~\ref{lem:supp} and Corollary~\ref{cor:suppress} that at most $\beta c(n)$ nodes can be hindered by the adversary.
Finally, we show in Lemmas~\ref{lem:UniversalInfluence} and \ref{lem:UniversalInfluence2} that there is a large set of nodes $V^*$ such that all nodes in $V^*$ are able to propagate their information to a large \emph{common} set of nodes.

In Sections~\ref{sec:support} and \ref{sec:supportAdaptive}, we describe how to use the previously derived techniques on information spreading to estimate the ``support'' (i.e.\ number) of nodes that belong to a specific category (either red or blue).
These protocols will form a fundamental building block for our \sa\ algorithms.

Due to the high amount of churn and the dynamically changing network, we use message flooding to disseminate and gather information. We now
precisely define flooding. Any node can initiate a  message for flooding.
Messages that need to be flooded have an indicator bit {\sc bFlood} set to 1.
Each of these messages also contains a terminating condition.  The initiating
node sends copies of the message to itself and its neighbors. When a node receives a
message with {\sc bFlood} set to 1, it continues  to send copies of that message to itself
and its neighbors in subsequent rounds until the terminating condition is
satisfied.

\subsection{Dynamic Distance and Influence Set} \label{sec:dynamic}
Informally, the dynamic distance from node $u$ to node $v$ 
is the number of rounds required for a message at $u$ to reach $v$.
We now formally define the notion of {\em dynamic distance} of a node $v$ from $u$ starting at
round $r$, denoted by $\dd_r(u \to v)$. When the subscript $r$ is omitted, we
 assume that $r=1$. 

Suppose node $u$ joins the network at round $r_u$, and,
from round $\max(r_u, r)$ onward, $u$ initiates a message $m$ for flooding whose
terminating condition is: {\sc $\msg{\text{has reached $v$}}$}. If $u$ is churned out before
$r$, then $\dd_r(u \to v)$ is undefined. Suppose  the first of those flooded
messages reaches $v$ in round $r+\Delta r$. Then, $\dd_r(u \to v) = \Delta r$.
Note that this definition allows $\dd_r(u \to v)$ to be infinite under two
scenarios. Firstly, node $v$ may be churned out before any copy of $m$ reaches
$v$. Secondly, at each round, $v$ can be shielded by churn nodes that absorb the
flooded messages and are then removed from the network before they can propagate
these messages any further. The influence set of a node $u$ after $R$ rounds
starting at round $r$ is given by: 
\begin{equation*}%
\infl_r(u, R) = \{v \in V^{r+R} : \dd_r(u \to v) \le R \}.
\end{equation*}
Note that we require  $\infl_r(u, R) \subseteq V^{r+R}$. Intuitively, we want
the influence set of $u$ (in this dynamic setting) to capture the nodes
\emph{currently} in the network that were influenced by $u$.
Note however  that the influence set of a node $u$ is meaningful even after $u$ is churned out.
Analoguously, we define 
\begin{equation*}\infl_r(U,R) = \cup_{u \in U} \infl_r(u,R),\end{equation*}
for any set of nodes $U\subseteq V^r$.

If we consider only a single node $u$, an (adaptive) adversary can easily prevent the
influence set of this node from ever reaching any significant size by simply
shielding $u$ with churn nodes that are replaced in every round.\footnote{An
oblivious adversary can achieve the same effect with constant probability for
linear churn.}

\subsection{Properties of Influence Sets}
We now focus our efforts on characterizing influence sets. This will help us in understanding how we can use flooding to spread information in the network.
For the most part of this section we assume that the network is controlled by an
adaptive adversary (cf.\ Section~\ref{sec:sa}).
The following lemma shows that the number of nodes that are sufficient to influence almost all the nodes in the network is given by the churn-expansion ratio (cf.\ Equation\ \eqref{eq:beta}):

\begin{lemma}\label{lem:beta}
Suppose that the adversary is adaptive. Consider any set $U \subseteq V^{r-1}$ (for
any $r \ge 1$) such that $|U| \ge \beta c(n)$.  
Then, after
\[
T = 2 \left \lceil \frac{\log n - \log c(n) - \log (\beta - \frac{\eps (1+\alpha)}{\alpha}) - 1 }{\log(1+\alpha) }\right \rceil
\] 
number of rounds, it holds that
\begin{equation}
\left|\infl_r(U, T)\right| > n- \beta c(n).
\end{equation}
When considering linear churn, i.e., $c(n)=n$, the bound $T$ becomes a constant
independent of $n$. On the other hand, when considering a churn order of
$\sqrt{n}$, we get $T \in O(\log n)$.
\end{lemma}
\longOnly{
\begin{proof}
Our proof assumes that $r=1$ for simplicity as the arguments extend quite easily to arbitrary values of $r$.
We proceed in two parts: First we show that the  nodes in $U$  influence at least $n/2$
nodes in some $T_1$ rounds. More precisely, we show that $|\infl(U,T_1)| \ge n/2$. We  use vertex expansion in a
straightforward manner to establish this part. Then, in the second part we show
that nodes in $\infl(U,T_1)$ go on to influence more than $n - \beta c(n)$ nodes.   We cannot use the vertex expansion in a straightforward manner in the second part because the cardinality of the set that is expanding in influence is larger than $n/2$. Rather, we use a slightly more subtle argument in which we use vertex expansion going backward in
time. The second part requires another $T_1$ rounds. Therefore, the two parts together complete the proof when we set $T = 2 T_1$. 

To begin the first part, consider $U \subseteq V^0$ at the start of round 1 with $|U| \ge \beta c(n)$. 
In round $1$, up to $\eps c(n)$ nodes in $U$ can be churned out. Subsequently, the remaining  nodes in $U$ influence some nodes outside $U$ as $G^1$ is
an expander with vertex expansion at least $\alpha$. More precisely, we can say that
\begin{equation}
|\infl(U,1)| \ge (\beta c(n) - \eps c(n)) (1+\alpha).
\end{equation} At the start of round $2$, the graph
changes dynamically to $G^{2}$. In particular, up to $\eps c(n)$ nodes  might be
churned out and they may all be in $\infl(U,1)$ in the worst case. However, the influenced set will again expand. Therefore, $|\infl(U,2)|$ cannot be less than
$(|\infl(U,1)| - \eps c(n))(1+\alpha) \ge \beta c(n)(1+\alpha)^2 - \eps c(n) (1+\alpha)^2 - \eps c(n) (1 +\alpha)$. Of course, there will be more churn at the start of round 3 followed by expansion leading to:
\begin{alignat*}{2}
|\infl(U,3)| 
  &\ge \Big(\beta c(n)(1+\alpha)^2 &-& \eps c(n) (1+\alpha)^2\\
  && -& \eps c(n) (1 +\alpha) \\
  && -& \eps c(n)\Big) (1+\alpha) \\
  &= \beta c(n) (1+\alpha)^3 &-& \eps c(n)\sum_{k=1}^3(1+\alpha)^k.
\end{alignat*}
This cycle of churn  followed by expansion continues and we get the following
bound at the end of some round $i$:
\begin{align*}
  |\infl(U,i)|  &\ge \beta c(n) (1+\alpha)^i - \eps c(n)\sum_{k=1}^{i}(1+\alpha)^k  \\
&= \beta c(n) (1+\alpha)^i \\
&+ \eps c(n) \frac{1- (1+\alpha)^{i+1}}{\alpha} - \eps c(n)
\end{align*}

Therefore, after 
\begin{equation}
T_1 = \left \lceil \frac{\log n - \log c(n) - \log (\beta - \frac{\eps
(1+\alpha)}{\alpha}) - 1 }{\log(1+\alpha)} \right \rceil\end{equation} 
rounds, we get 
\begin{equation} \label{eq:part1}
  |\infl(U,T_1)|\ge n/2\text{.}
\end{equation}

Now we move on to the second part of the proof. Let $T = 2 T_1$.
If $|\infl(U,T)| > n - \beta c(n)$, we are done.
Therefore, for the sake of a contradiction,  assume that  $|\infl(U,T)| \le 
n-\beta  c(n)$. Let $S = V^{T} \setminus \infl(U,T)$, i.e., $S$ is the set of nodes in $V^T$ that were not influenced by $U$ at (or before) round $T$. 
Moreover, $|S| \ge \beta c(n)$ because we have assumed that $|\infl(U,T)| \le
n-\beta  c(n)$. We will start at round $T$ and work our way
backward. For $q \le T$, let $S^q \subseteq V^{q}$, be the set of
all vertices in $V^q$ that, starting from round $q$, influenced  some vertex in $S$ at or
before round $T$. More precisely, \[S^q = \set{s \in V^{q} : 
\infl_{q}(s, T - q) \cap S \ne \emptyset}.\]
Suppose that $|S^{T_1}| > n/2$. Then 
  \begin{equation*}S^{T_1} \cap \infl(U,T_1) \ne \emptyset\text{,}\end{equation*}
since $|\infl(U,T_1)| \ge n/2$ by
\eqref{eq:part1}. Consider a node
$s^* \in  S^{T_1} \cap \infl(U,T_1)$. Note that $s^*$ was influenced by $U$ and went
on to influence some node in $S$ before (or at) round $T$. However, by definition, no node  in $S$ can be influenced by any node in $U$ at or before round $T$. We have thus reached a contradiction. 

We are left with showing that $|S^{T_1}| > n/2$.
We start with $S$ and work our way backwards. We know that $|S| \ge \beta c(n) >
\beta c(n) - \eps c(n)$. We want to compute the cardinality of $S^{T-1}$. We
first focus on an intermediate set $S'$, which we define as 
\begin{equation*}S'= S \cup \set{s': \exists (s,s') \in E^T}.\end{equation*} 
Since $G^T$ is an expander, $|S'| \ge |S|(1+\alpha)$. 
Furthermore, it is also clear that each node in $S'$ could influence some node in $S$. 
Notice that $ S' \setminus S^{T-1}$ is the set of nodes in $S'$ that were churned in only at the start of round $T$. 
Therefore, 
\begin{align*}
|S^{T-1}| &\ge |S'| - \eps c(n) \\
&\ge |S|(1+\alpha) - \eps c(n) \\
&> (\beta c(n) - \eps c(n))(1+\alpha) - \eps c(n) \\
&= \beta c(n)(1+\alpha) - \eps c(n)(1+\alpha) - \eps c(n).
\end{align*}
Continuing to work our way backwards in time, we get
\begin{align*}
|S^{T-2}| > \beta c(n)(1+\alpha)^2 &- \eps c(n)(1+\alpha)^2\\ 
                                   &- \eps c(n)(1+\alpha) - \eps c(n),
\end{align*}
Or more generally,
\begin{align*}
|S^{T-i}| &> \beta c(n)(1+\alpha)^i - \eps c(n) \sum_{0 \le j \le i} (1+\alpha)^j \\
&= \beta c(n)(1+\alpha)^i + \eps c(n) \frac{1 - (1+\alpha)^{i+1}}{\alpha}\\
&= \beta c(n)(1+\alpha)^i  - \frac{\eps c(n) (1+\alpha)^{i+1}}{\alpha} + \frac{\eps c(n)}{\alpha}.
\end{align*}
We now want the value of $i$ for which \begin{equation*}|S^{T-i}|  > n/2 + \frac{\eps
c(n)}{\alpha} > n/2.\end{equation*} In other words, we want a value of $i$ such that
\[
\beta c(n)(1+\alpha)^i  - \frac{\eps c(n) (1+\alpha)^{i+1}}{\alpha} + \frac{\eps c(n)}{\alpha} > n/2 + \frac{\eps c(n)}{\alpha},
\]
which is obtained when $i = T_1$. Therefore, it is
easy to see that if we set $T = 2 T_1 $, we get $|S^{T_1}|
> n/2$, thereby completing the proof. 
\qed
\end{proof}
}
At  first glance, it might appear to be counterintuitive that the order of the
bound $T$ decreases with increasing churn. When the adversary has the benefit of churn that is linear in $n$, our bound on $T$ is a constant, but when the adversary is limited to a churn order of $\sqrt{n}$, we get 
$T \in O(\log n)$. This, however, turns out to be fairly
natural when we note that the size of the set $U$ of nodes that we start out with is in proportion to the churn limit.

We say that a node $u \in V^r$ is \emph{suppressed for $R$ rounds} or \emph{shielded by churn} if $|\infl_r(u, R)|
< n - \beta c(n)$; otherwise we say it is \emph{unsuppressed}. The following lemma shows
that given a set with cardinality at least $\beta c(n)$ some node in that set will be
unsuppressed. 

\begin{lemma}\label{lem:supp}
Consider the adaptive adversary.
Let $U$ be any subset of $V^{r-1}$, $r \ge 1$, such that $|U| \ge \beta c(n)$.
Let $T$ be the bound derived in
Lemma~\ref{lem:beta}. There is at least one $u^* \in U$ such that for some $R
\in O(T\log n)$, $u^*$ is unsuppressed, i.e.,  
\[|\infl_r(u^*,R)| > n- \beta c(n).\]
In particular, when the
order of the churn is $n$, $T$ becomes a constant, and we have $R = O(\log n)$. 
\end{lemma}
\onlyLong{
Before we proceed with our key arguments of the proof, we state a property of bipartite graphs that we will use subsequently.
\begin{property} \label{prop:bipartite}
Let $H=(A,B,E)$ be a bipartite graph in which $|A| > 1$ and every vertex $b \in B$ has at least one neighbor in $A$. There is a subset $A^* \subset A$ of cardinality at most 
$\lceil |A|/2 \rceil$ such that  
\begin{equation*}|\set{b : \exists a^* \in A^* \text{such that } (a^*, b) \in E}| \ge \lceil
|B|/2 \rceil.\end{equation*}
\end{property}
\begin{proof}(of Property~\ref{prop:bipartite})
Consider each node in $A$ to be a unique color. Color each node in $B$ using the color of a neighbor in $A$ chosen arbitrarily. Now partition $B$ into maximal subsets of nodes with like colors. 
Consider the parts of the partition sorted in decreasing order of their cardinalities. 
We  now greedily choose the first  $\lceil |A|/2 \rceil$ colors in the sorted order of parts of $B$. We call the chosen colors $C$. Observe that colors in $C$ cover at least as many nodes in $B$ as those not in $C$.
Suppose the colors in $C$  cover fewer than $\lceil |B|/2 \rceil$ nodes in $B$.
Then the remaining colors will cover $\lceil |B|/2 \rceil$, but that is a
contradiction. Therefore, colors in $C$ cover at least $\lceil |B|/2 \rceil$
nodes in $B$. The nodes in $A$ that have the colors in $C$ are the nodes that
comprise $A^*$, thereby completing our proof. \qed
\end{proof}
\begin{proof}(of Lemma~\ref{lem:supp})
Again, our proof assumes $r =1$ because it generalizes to arbitrary values of
$r$ quite easily.  From Lemma~\ref{lem:beta}, we know that the influence of all
nodes in $U$ taken together will reach $n - \beta c(n)$ nodes in $T$ rounds.
This does not suffice because we are interested in showing that there is at least one
node in $V^0$ that (individually) influences $n- \beta c(n)$ nodes in $V^R$ for some $R = {O(T\log n)}$.

From Lemma~\ref{lem:beta}, we know that $U$ (collectively) will influence at least $n - \beta
c(n)$ nodes in T rounds, i.e.,
\begin{equation*}|\infl(U,T)| > n- \beta c(n)\text{.}\end{equation*}
From Property~\ref{prop:bipartite}, we know that there is a set $U_1 \subset U$
of cardinality at most $\lceil |U|/2 \rceil$ such that \[|\infl(U_1,T)| >
\frac{n- \beta c(n)}{2}.\]
Recalling
that $\beta < \frac{1}{12} < \frac{1}{3}$, we know that $|\infl(U_1,T)| \ge \beta
c(n)$.  We can again use Lemma~\ref{lem:beta} to say that $\infl(U_1,T)$ influences
more than $n- \beta c(n)$ nodes in additional $T$ rounds and, by transitivity,
$U_1$ influences more than $n- \beta c(n)$ nodes after $2T$ rounds. We therefore
have $|\infl(U_1,2T)| > n- \beta c(n)$. Again, we can choose a set $U_2\subset
U_1$ (using Property~\ref{prop:bipartite}) that consists of $\lceil|U_1|/2\rceil$ nodes in $U_1$ such  that $|\infl(U_2,2T)| \ge \beta c(n)$. Subsequently applying
Lemma~\ref{lem:beta} extends the influence set of $U_2$ to more than $n- \beta c(n)$ after
$3T$ rounds. 

In every iteration $i$ of the above argument, the size of the set $U_i$ decreases
by a constant fraction until we are left with a single node
$u^* \in U$ such that $|\infl(u^*, O(\log n) T)| > n- \beta c(n)$.
\qed
\end{proof}
}

Can $\beta c(n)$ (or more nodes)  be suppressed for any significant number of (say, $\Omega(T \log n)$) rounds? This is in immediate contradiction to Lemma~\ref{lem:supp} because any such suppressed set of nodes must contain an unsuppressed node. This leads us to the following corollary.

\begin{corollary}\label{cor:suppress}
The number of nodes that can be suppressed for $\Omega(T \log n)$ rounds is less
than $\beta c(n)$, even if the network is controlled by an adaptive
adversary.
\end{corollary}
\onlyLong{
\begin{corollary}\label{cor:random}
Consider an oblivious adversary that must commit to the entire
sequence of graphs in advance. If we choose a node $u$ uniformly at random
from $V^0$, with probability at least $1 - \frac{\beta c(n)}{n}$,
then $u$ will be unsuppressed, i.e.,
\begin{equation*}\left|\infl(u,
\Omega(T\log n))\right| > n-\beta c(n).\end{equation*}
\end{corollary}
\begin{proof}
Let $S \subset V^0$ be the set of nodes suppressed for $\Omega(T\log n)$ rounds. Under an oblivious adversary, the node $u$ chosen unformly at random from $V^0$ will not be in $S$ with probability $1 - \frac{\beta c(n)}{n}$, and hence, will not be suppressed with that same probability.
\qed
\end{proof}

The following two lemmas show that there exists a set $V^*$ of unsuppressed nodes, all of which can influence a large common set of nodes, given enough time.

\begin{lemma}\label{lem:UniversalInfluence}
Consider a dynamic network under linear churn that is controlled by an adaptive
adversary. In some $r \in O(\log n)$ rounds, there
is a set of unsuppressed nodes $V^* \subseteq V^0$ of cardinality more than $(1 -
\beta)n$ such that \[\left | \bigcap_{v \in V^*} \infl(v, r) \right | >
(1-\beta)n.\] 
\end{lemma}

\begin{proof}
Let $V^* \subseteq V^0$ be any set of unsuppressed nodes, i.e., in some $c_0
\log n$ rounds for some constant $c_0$, the influence set of each $v \in V^*$
has cardinality more than $(1 - \beta) n$. Note that, however, we {\em cannot}
guarantee that, for any two vertices $v_1$ and $v_2$ in $V^*$, 
\begin{equation*}|\infl(v_1, c_0 \log n) \cap \infl(v_1, c_0 \log n)| > (1 - \beta) n.\end{equation*} 
Assume for simplicity
that $|V^*|$ is a power of 2. Consider any pair of vertices $\{v_1, v_2\}$, both
members of $V^*$. Recalling that $\beta < \frac{1}{12} < \frac{1}{3}$, we can say that
\begin{equation*}\left|\infl(v_1, c_0 \log n) \cap \infl(v_2, c_0 \log n)\right| \ge \beta n.\end{equation*} Therefore,
considering that the intersected set $\infl(v_1, c_0 \log n) \cap \infl(v_2, c_0 \log
n)$ of nodes has cardinality at least $\beta n$, we can apply
Lemma~\ref{lem:beta} leading to $|\infl(v_1, c_0 \log n + T) \cap \infl(v_2, c_0
\log n+T)| > (1-\beta) n$.  We can partition $V^*$ into  a set $S_1$ of
$\frac{|V^*|}{2}$ pairs such that for each pair, the intersection of influence
sets has cardinality more than $(1-\beta)n$ after $c_0 \log n+T$ rounds.
Similarly, we can construct a set $S_2$ of quadruples by disjointly pairing the
pairs in $S_1$. Using a similar argument, we can say that for any  $Q\in S_2$,
\begin{equation*}\left|\bigcap_{v \in Q} \infl(v, c_0 \log n + 2T)\right| > (1-\beta) n.\end{equation*}
Progressing
analogously, the set $S_{\log |V^*|}$ will equal $V^*$ and we can conclude that 
\[
\left |\bigcap_{v \in S_{\log |V^*|}} \infl(v, c_0  \log n + T \log |V^*|) \right | > (1-\beta) n.
\]
Since $|V^*| \le n$, it holds that $c_0  \log n + T \log |V^*| \in O(\log n)$, thus completing the proof. 
\qed
\end{proof}
}

\onlyLong{
\begin{lemma}\label{lem:UniversalInfluence2}
Suppose that up to $\eps\sqrt{n}$ nodes can be subjected to churn in any round
by an adaptive adversary. In
some $r \in O(\log^2 n)$ rounds, there is a set of unsuppressed nodes $V^*
\subseteq V^0$ of cardinality at least $n - \beta \sqrt{n}$ such that \[\left |
\bigcap_{v \in V^*} \infl(v, r) \right | > n -\beta \sqrt{n}.\] 
\end{lemma}
\begin{proof}
Since we assume that $c(n)=\sqrt{n}$, the bound $T$ of Lemma~\ref{lem:beta} is
in $O(\log n)$.
Therefore, by instantiating Corollary~\ref{cor:suppress}, we know that each of the unsuppressed nodes
in $V^*$ (which is of cardinality at least $n - \beta \sqrt{n}$) will influence
more than $n - \beta \sqrt{n}$ nodes in $O(\log^2 n)$ time. We can use  the same
argument as in Lemma~\ref{lem:UniversalInfluence} to show that in $O(\log n)$
rounds, all the unsuppressed nodes have a common influence set of size at least
$\Theta(n)$. That common influence set will grow to at least $n - \beta \sqrt{n}$
nodes within another $O(\log^2 n)$ rounds. Thus a total of $O(\log^2 n)$ rounds
is sufficient to fulfill the requirements.
\qed
\end{proof}
}

\shortOnly{
In a dynamic network with churn limit $\eps n$, the entire set of nodes in the
network can be churned out and new nodes churned in within $1/\eps$ rounds. This
calls for maintaining important global information (such as a global clock) in
the network. This can be achieved quite easily by appropriate message flooding.
Details can be found in the full paper \cite{APRU11}.
}
\longOnly{
\subsection{Maintaining Information in the Network} \label{subsec:maintain}
In a dynamic network with churn limit $\eps n$, the entire set of nodes in the
network can be churned out and new nodes churned in within $1/\eps$ rounds. How
do the new nodes even know what algorithm is running? How do they know how far
the algorithm has progressed? To address these basic questions, the network
needs to maintain some global information that is not lost as the nodes in the
network are churned out. There are two basic pieces of information that need to
be maintained so that a new node can join in and participate in the execution of
the distributed algorithm: 
\begin{enumerate}
	\item the algorithm that is currently executing, and
	\item the number of rounds that have elapsed in the execution of the algorithm. In other words, a global clock has to be maintained.
\end{enumerate}
We assume that the nodes in $V^0$ are all synchronized in their understanding of what algorithm to execute and the global clock. The nodes in the network continuously flood information on what algorithm is running so that when a new node arrives, unless it is shielded by churn, it receives this information and can start participating in the algorithm. To maintain  the clock value, nodes send their current clock value to their immediate neighbors. When a new node receives the clock information from a neighbor, it sets its own clock accordingly. Since  nodes are not malicious or faulty,  Lemma~\ref{lem:beta} ensures that information is correctly maintained in more than $n - \beta c(n)$ nodes.}

\subsection{Support Estimation Under an Oblivious Adversary } \label{sec:support} \label{sec:supportOblivious}
Suppose we have a dynamic network  with $\R$ nodes colored red in $V^0$. $\R$ is also called the {\em support} of red nodes. 
We want the nodes in the network to estimate $\R$ under an oblivious adversary.
We assume that the adversary chooses $\R$ and which $\R$ nodes in $V^0$ to color red, but it
does not know the random choices made by the algorithm. Furthermore, we assume that churn can be linear in $n$, i.e., $c(n) = n$. 

Our algorithm uses random numbers drawn from the exponential distribution, whose probability density function, we recall, is parameterized by $\lambda$ and given by $f(x) = \lambda \exp (- \lambda x)$ for all $x \ge 0$. Furthermore, we notice that the expected value of a random number drawn from the exponential distribution of parameter $\lambda$ is $1/\lambda$.
\onlyLong{
We now present two properties of exponential random variables that are crucial to our context. Consider $K\ge 1$  independent random variables $Y_1, Y_2, \ldots, Y_K$, each following the exponential distribution of rate $\lambda$. 
\begin{property}[see~\cite{grinstead1997introduction} for example]
The minimum among all $Y_i$'s, for $1 \le i \le K$, is  an exponentially distributed random variable with parameter $K \lambda$. \label{prop:min}
\end{property}
The idea behind our algorithm exploits Property~\ref{prop:min} in the following manner. If each of the $\R$ red nodes  generate an exponentially distributed random number with parameter 1, then the minimum $\bar{s}$ among those $\R$ random numbers will also be  exponentially distributed, but with parameter $\R$. Thus $1/\bar{s}$ serves as an estimate of $\R$. To get a more accurate estimation of $\R$, we exploit the following property that provides us with sharp concentration when the process is repeated a sufficient number of times.
\begin{property}[see \cite{AoyamaS08} and pp.\ 30, 35 of \cite{DemboZ98}]  \label{prop:shah}
Let $X_K = \frac{1}{K} \sum_{i=1}^K Y_i$. Then, for any $\varsigma \in (0, 1/2)$, 
\[
Pr\left ( \left | X_K - \frac{1}{\lambda} \right | \ge \frac{\varsigma}{\lambda}\right ) \le 2 \exp \left( -\frac{\varsigma^2 K}{3}\right).
\]
\end{property}
}

\begin{algorithm}[h]
  \begin{algorithmic}[1]
   \footnotesize
  \EMPTY The following pseudocode is executed at every node $u$.
  
  \EMPTY $P  \in \Theta(\log n)$ controls the precision of our estimate. Its exact value is worked
out in the proof of Theorem~\ref{thm:ObliviousSupport}.
 
   \EMPTY 
  \item[\bf At round 1:]
  
  \STATE  Draw $P$ random numbers $s_1, s_2, \ldots,
s_i, \ldots,  s_{P}$, each from the exponential random distribution with rate
$1$. 

\COMMENT{Each $s_i$ is chosen with a precision that ensures that the smallest possible positive value is at most $\frac{1}{n^{\Theta(1)}}$;} 

\COMMENT{Note that $\Theta(\log n)$ bits suffice.}

\STATE For each $s_i$, create a message $m_u(i)$ containing $s_i$ and a terminating condition: {\sc has encountered
a message $m_v(i)$ with a smaller random number}.

\COMMENT{Notice that a node $u$ will flood exactly one message at each index $i$ --- in particular the  smallest random number encountered by node $u$ with message index $i$}

\STATE For each $i$, initiate flooding of message $m_u(i)$.

   \EMPTY 
  \item[\bf For the next $t = \Theta(\log n)$ rounds:]
  
\STATE Continue flooding messages respecting their termination conditions.

\COMMENT It is easy to see that the number of bits transmitted per round through a link is at most $O(\log^2 n)$.

 \EMPTY 
  \item[\bf At the end of the $\Theta(\log n)$ rounds:]
  
   \STATE For each $i$, the node $u$ holds a message $m_v(i)$. Let $\bar{s}_u(i)$ be the random number contained in $m_v(i)$. 
   \STATE $\bar{s}_u \leftarrow \frac{\sum_i \bar{s}_u(i)}{P}$. \label{lno:bars}
   \STATE Node $u$ outputs $1/\bar{s}_u$ as its estimate of $\R$.
   \COMMENT{Now that the estimation is completed, all messages can be terminated.}
    \end{algorithmic}
  \caption{Algorithm to estimate the support $\R$ of red nodes when $\R \ge n/2$.}\label{alg:support}
\end{algorithm}
\normalsize

We now present our algorithm for estimating $\R$ in pseudocode format (assuming $\R \ge n/2$); see Algorithm~\ref{alg:support}.

\begin{theorem}\label{thm:ObliviousSupport}
Consider an oblivious adversary and let $\gamma$ be a an arbitrary fixed constant $\ge 1$.
Let $\bar{\R} = \max(\R, n-\R)$. By executing Algorithm~\ref{alg:support} to estimate both $\R$ and $n-\R$, we can estimate $\bar{\R}$ to within $[(1 - \delta) \bar{\R}, (1 + \delta) \bar{\R}]$ for any  $\delta > 2 \beta$ with probability at least $1-n^{-\gamma}$.
\end{theorem}
\onlyLong{
\begin{proof} 

Without loss of generality, let $\R \ge n/2$. 
Out of the $\R$ red nodes up to $\beta n$ nodes (chosen obliviously) can be
suppressed, leaving us with 
\begin{equation}
\R' \ge \R - \beta n \ge (1-2\beta) \R \label{eqn:R'}
\end{equation}
unsuppressed red nodes (since $\R \ge n/2$). In a slight abuse of notation, we
use $\R$ and $\R'$ to denote both the cardinality and the set of  red nodes and
unsuppressed red nodes, respectively.  We define \[U = \bigcap_{v \in \R'} \infl(v, t);\] 
note that $t=O(\log n)$ and
$|U| \ge (1-\beta)n$ (cf.\ Lemma~\ref{lem:UniversalInfluence}). Let $u$
be some node in $U$. Let \[V_u = \set{v: v \in \R \wedge u \in \infl(v,t)}.\]
For all $u \in U$, $\R' \subseteq V_u \subseteq \R$. 
Notice that $\bar{s}_u(i)$ computed by $u$ in line number~\ref{lno:bars} of Algorithm~\ref{alg:support} is based on random numbers generated by all nodes in $V_u$.
Therefore, at round $t$,
node $u$ is estimating $\R$ using the exponential random numbers that  were
drawn by nodes in $V_u$. Since our adversary is oblivious, the choice of $V_u$
is independent of the choice of the random numbers generated by each $v\in V_u$.
Therefore, $\bar{s}_u(i)$ is an exponentially distributed random number with
rate  $|V_u| \ge \R'$ (cf.\ Property~\ref{prop:min}). For any $\delta  > 2\beta$,
let $\varsigma \le \min \set{\frac{\delta  - 2\beta}{1-\delta }, \frac{\delta
}{1+\delta }}$. When $P = \frac{3 \gamma \ln n}{\varsigma^2} \in \Theta(\log n)$ parallel
iterations are performed, where $\gamma\ge 1$, the required accuracy is obtained with
probability $1 - \frac{1}{\Omega(n^{\gamma})}$ (cf.\ Property~\ref{prop:shah}). 
\qed
\end{proof}

\subsection{Support Estimation Under an Adaptive Adversary} \label{sec:supportAdaptive}
The algorithm for support estimation under an oblivious adversary (cf.\ 
Section~\ref{sec:supportOblivious}) does not work under an adaptive 
adversary. To
estimate the support of red nodes in the network, each red node draws
a random number from the exponential distribution and floods it in an attempt to
spread the smallest random number. When the adversary is adaptive, the smallest
random numbers can easily be targeted and suppressed. To mitigate this difficulty, we consider a different algorithm in which the number of bits communicated is larger. In particular, the number of bits communicated per round by each node executing  this algorithm is at most polynomial in $n$.

Let $\R$ be the
support of the red nodes. Every node floods its unique identifier along with a
bit that indicates whether it is a red node or not.  At most $\beta \sqrt{n}$
nodes' identifiers can be suppressed by the adversary for $\Omega(\log^2 n)$ rounds leaving  at least $n-\beta
\sqrt{n}$ unsuppressed identifiers (cf.\ Corollary~\ref{cor:suppress}). Each node
counts the number of unique red identifiers $A$ and non-red identifiers $B$ that
flood over it and estimates $\R$ to be $A + \frac{n-A-B}{2}$.  

This support
estimation technique generalizes quite easily to arbitrary churn order.
Therefore, we state the following theorem more generally.
\begin{theorem}\label{thm:AdaptiveSupport}
Consider the algorithm mentioned above in which nodes flood their unique
identifiers indicating whether they are red nodes or not and assume that the
network is controlled by an adaptive adversary.  Let $c(n)$ be the
order of the churn; we assume for simplicity that $c(n)$ is either $n$ or
$\sqrt{n}$.  Then the following holds:
\begin{compactenum}
	\item At least $n-\beta c(n)$  nodes estimate $\R$  between $\R - \frac{\beta c(n)}{2}$ and $\R+ \frac{\beta c(n)}{2}$. Furthermore, these nodes are aware that their estimate is within $\R - \frac{\beta c(n)}{2}$ and $\R+ \frac{\beta c(n)}{2}$.
	\item The remaining nodes are aware that their estimate of $\R$ might fall
    outside $[\R - \frac{\beta c(n)}{2}, \R+ \frac{\beta c(n)}{2}]$.
	\end{compactenum}
	When $c(n) = n$, it requires only $O(\log n)$ rounds, but when $c(n)=\sqrt{n}$, it requires $O(\log^2 n)$ rounds.
\end{theorem}

\begin{proof}
Let $u$ be any one of the $n-\beta c(n)$ nodes that receive at least $n-\beta
c(n)$ unsuppressed identifiers (cf.\ Lemma~\ref{lem:UniversalInfluence} and
Lemma~\ref{lem:UniversalInfluence2}). Let $A$ and $B$ be the number of unique
identifiers from red nodes and non-red nodes, respectively, that flood over $u$.
Let $C = n - A - B \le \beta c(n)$. This means that $u$ estimates $\R$ to be
$A+\frac{C}{2}$. Note that $ A \le \R \le A+C$ and since $C \le \beta c(n)$, $\R$
is estimated between $\R - \frac{\beta c(n)}{2}$ and $\R+ \frac{\beta c(n)}{2}$.
Furthermore, since $u$ received $n - \beta c(n)$ identifiers, it can be sure
that its estimate is between $\R - \frac{\beta c(n)}{2}$ and $\R+ \frac{\beta
c(n)}{2}$.

If a node does not receive at least $n - \beta c(n)$ identifiers, then it is
aware that its estimate of $\R$ might not be within $[\R - \frac{\beta
c(n)}{2}, \R+ \frac{\beta c(n)}{2}]$.

From Lemma~\ref{lem:UniversalInfluence}, when $c(n) = n$,  the algorithm takes $O(\log n)$ rounds to complete because we want to ensure that unsuppressed nodes have flooded the network. When $c(n) = \sqrt{n}$, as a consequence of Lemma~\ref{lem:UniversalInfluence2}, the algorithm requires $O(\log^2 n)$ rounds. 
\qed
\end{proof}

}

\section{\sa\ Under an Oblivious Adversary} \label{sec:oblivious}

In this section we will first present Algorithm~\ref{alg:bc} for the simpler
problem of reaching \bc, where the input values are restricted to $\set{0,1}$
(cf.\ Section~\ref{sec:sa}). We will then use this algorithm as a subroutine
for solving \sa\ in Section~\ref{sec:obliviousSA}.

Throughout this section we assume suitable choices of $\eps$ and $\alpha$ such
that the upper bound 
\begin{equation} \label{eq:betaUpperOblivous}
\beta<\frac{1}{12}
\end{equation}
can be satisfied for $\beta$; note that \eqref{eq:betaUpperOblivous} must hold
in addition to bound \eqref{eq:beta}.
Moreover, we assume that a node can send and process up to
$O(\log^2 m)$ bits in every round, where $m$ is the size of the input value
domain.  

\subsection{\bc} \label{sec:obliviousBC}

A node $u$ that executes Algorithm~\ref{alg:bc} proceeds in a sequence of
$O(\log n)$ checkpoints that are interleaved by $O(\log n)$ rounds. Each node $u$ has a bit variable $b_u$ that stores its current output value. At each
checkpoint $t_i$, node $u$ initiates support estimation of the number of nodes
currently having 1 as their output bit by using the algorithm described in
Section~\ref{sec:support}. (At checkpoint $t_{R-1}$, nodes estimate both:
the support of 1 and 0.) 
The outcome of this support estimation will be
available in checkpoint $t_{i+1}$ where $u$ has derived the estimation $\#(1)$.
If $u$ believes that the support of 1 is small ($\le \frac{1}{4}n$), it
sets its own output $b_u$ to 0; if, on the other hand, $\#(1)$ is large ($\ge
\frac{3}{4}n$), $u$ sets its output $b_u$ to 1. This guarantees stability once
agreement has been reached by a large number of nodes. When the support of 1 is
roughly the same as the support of 0, we need a way to sway the decision to one
side or the other. This is done by flooding the network whereby the flooding
message of node $v$ is weighted by some randomly chosen value, say $r_v$. The adversary can only
guess which node has the highest weight and therefore, with constant
probability, the flooding message with this highest weight (i.e., smallest
random number) will be used to set the output bit by almost all nodes in the
network.

\begin{algorithm}[h]
  \begin{algorithmic}[1]
  \footnotesize
  \EMPTY Let $decision_u$ be initialized to $\bot$.
  \EMPTY
  Let $b_u$ be the current output bit of $u$. If $u \in V^0$, then $b_u$ is
  initialized to the input value of $u$; otherwise it is set to $\bot$. 
  
  Let $t_1 = 1$ be the first checkpoint round. Subsequent checkpoint rounds are
  given by  $t_i = t_{i-1} + O(\log n)$, for $i > 1$. 
  For the terminating checkpoint $t_R$, we choose an $R \in O(\log n)$, i.e.,
  $t_R\in O(\log^2n)$.
  \EMPTY 
  \item[\bf At every checkpoint round $t_i$ \underline{excluding} $t_R$:] 
    
  \STATE Initiate support estimation (to be completed in checkpoint round $t_{i+1}$).
  \STATE Generate a random number $r_u$ uniformly from $\set{1, \ldots, n^k}$ for suitably large but constant $k$. (With high probability, we want exactly one node to have generated $\min_u r_u$.)
  \STATE Initiate flooding of $\set{r_u, b_u}$ with terminating condition: {\sc
  $\langle($has encountered another message initiated by $v \ne u$ with $r_v <
  r_u$) $\vee$ (current round $\ge t_{i+1}$)$\rangle$.}
  
  \EMPTY
  \item[\bf At every checkpoint round $t_i$ \underline{except} $t_1$:] 
  
  \STATE Use the support estimation initiated at checkpoint round $t_{i-1}$. Let
  $\#(1)$ be $u$'s estimated support value for the number of nodes that had an
  output of 1. 
  \IF{$\#(1) \leq  \frac{1}{4}n$} \label{line:onequarter}
    \STATE $b_u \leftarrow 0$.
  \ELSIF{$\#(1) \geq \frac{3}{4}n$}
    \STATE $b_u \leftarrow 1$.
  \ELSIF{$u$ has received flooded messages initiated in $t_{i-1}$}
    \STATE Let $\set{r_v, b_v}$ be the message with the smallest random number that flooded over $u$.
    \STATE $b_u \leftarrow b_v$.
  \ENDIF
  
  \EMPTY
  \item[\bf At terminating checkpoint round $t_R$:]
  \IF{$\#(1) \ge  \frac{n}{2}$} \label{line:oneHighOblivious}
   		\STATE $decision_u \la 1$.
      \STATE Flood a 1-decision message ad infinitum.
  \ELSIF{$\#(0) \ge \frac{n}{2}$}
   		\STATE $decision_u \la 0$.
      \STATE Flood 0-decision message ad infinitum.
  \ENDIF
  \EMPTY
  \item[\bf If $u$ receives a $b$-decision message:]
    \STATE $decision_u \la b$
  \end{algorithmic}
  \caption{\bc\ under an oblivious adversary; code executed by node $u$.} \label{alg:bc}
\end{algorithm}
\normalsize

\begin{theorem} \label{thm:ssd}
Assume that the adversary is oblivious and that the churn limit per round is
$\eps n$.
Algorithm \ref{alg:bc} %
reaches stability in $O(\log^2 n)$ rounds and achieves \bc\ with high probability.
\end{theorem}
\onlyLong{
\begin{proof} 
  Throughout this proof we repeatedly invoke the properties of the support estimation as stated in Theorem~\ref{thm:ObliviousSupport}, which succeeds with probability $1-1/n^{\gamma}$.
  Assuming that $\gamma \ge 2$, suffices to guarantee that all of the $\Theta(\log n)$ invocations of the support estimation are accurate with high probability.

We first argue that Validity holds: Suppose that all nodes start with
input value~1. The only way a node can set its output to 0 is by passing
Line~\ref{line:onequarter}. This can happen for at most $\beta n$ nodes.
The only way that more nodes can set their output to 0 is if they estimate the
support of 1 to be in $(\frac{1}{4}n,\frac{3}{4}n)$.  If $\beta$ is suitably
small, Theorem~\ref{thm:ObliviousSupport} guarantees that with high probability this
will not happen at any node. 
The argument is analogous for the case where all nodes start with 0.

Next we show Almost Everywhere Agreement:
Let $N_{i}$ be the number of nodes at checkpoint round $t_{i}$ that output 1.
Let $\lo_i$, $\hi_i$, and $\mid_i$, respectively, be the sets of nodes in
$V^{t_i}$ for which $\#(1) \leq  \frac{1}{4}n$, $\#(1) \geq  \frac{3}{4}n$, and 
$\frac{1}{4}n < \#(1) < \frac{3}{4}n$; note that nodes are placed in $\lo_i$,
$\hi_i$, and $\mid_i$ based on their $\#(1)$ values, which are estimates of
$N_{i-1}$, not $N_i$. Clearly, we have that $\lo_i + \mid_i + \hi_i = n$.

For some $i>1$, let $u^* \in V^{t_{i-1}}$ be the node that generated the
smallest random number in checkpoint round $t_{i-1}$ among all nodes in
$V^{t_{i-1}}$. With high probability, $u^*$ will be unique. By
Corollary~\ref{cor:random}, with probability $1-\beta$ (a constant), $u^*$ is
unsuppressed, implying that $b_{u^*}$ will be used by all nodes in $\mid_i$.
Consider the following cases:
\begin{description}
\item[Case A ($N_{i-1} \le (\frac{1}{4} - \delta)n$):] From
  Theorem~\ref{thm:ObliviousSupport}, we know that with high probability
  $|\lo_i| \ge (1-\beta)n$ implying $|\mid_i| +|\hi_i| \le \beta n$.
  Therefore, $N_{i}$ will continue to be very small leading to small estimates
  $\#(1)$ in subsequent checkpoints. After $O(\log n)$ checkpoints, this causes
  at least $(1-\beta)n$ nodes to decide on $0$, with high probability. Moreover,
  it is easy to see that the remaining $\beta n$ nodes will not be able to pass
  Line~\ref{line:oneHighOblivious}, since the adversary cannot artificially
  increase the estimated support of nodes with 1. (Recall from
  Section~\ref{sec:support} that by suppressing
  the minimum random variables, the adversary can only make the estimate
  smaller.)
  
  (We are presenting separate Cases B, C, and D for clarity. Equivalently, we could have treated them together as one case with the condition that  $(\frac{1}{4} - \delta)n < N_{i-1} <(\frac{3}{4} + \delta)n$ leading to the implication that with high probability either $|\lo_i| +|\mid_i| \ge (1-\beta)n$ or $|\hi_i| +|\mid_i| \ge (1-\beta)n$.)

\item[Case B ($\frac{1}{4} - \delta)n < N_{i-1} < (\frac{1}{4} + \delta) n$):] With high
   probability, $|\lo_i| +|\mid_i| \ge (1-\beta)n$ implying  $|\hi_i| \le
   \beta n$. Note first that nodes in $\lo_i$ will set their output bits to 0.
   Since $N_{i-1} < (\frac{1}{4} + \delta) n$, there are at least $(\frac{3}{4}
   - \delta) n$ nodes
   in $V^{t-1}$ that output 0. Of these, at most $\beta n$ could have been
   suppressed. So, with probability at least $\frac{3}{4} - \delta- \beta$, $u^*$ is an
   unsuppressed node that outputs 0. When $u^*$ outputs 0, nodes in $\mid_i$
   will set their output bits to 0. Thus, considering $\lo_i$ and $\mid_i$, we
   have at least $(1-\beta)n$ nodes that set their output bits to $0$ with
   constant probability. 
For a suitably small $\delta$ and $\beta<\frac{1}{4}-\delta$, this will lead to
Case A in the next iteration, which means that subsequently nodes agree on 0.
	 
   \item[Case C ($(\frac{1}{4} +\delta) n \le N_{i-1} \le (\frac{3}{4} - \delta) n $):]
     With high probability, $|\mid_i| \ge (1-\beta)n$. With constant probability
     $(1-\beta)$, $u^*$ will be an unsuppressed node and nodes in $\mid_i$ will
     set their output bits to the same value $b_{u^*}$. This will  lead to 
     Case A in the next iteration. 

   \item[Case D ($(\frac{3}{4} - \delta)n < N_{i-1} < (\frac{3}{4} + \delta)n$):] 
     This is similar to Case B, i.e., with constant probability, at least $(1-
     \beta)n$  nodes will reach agreement on 1.

   \item[Case E ($N_{i-1} \ge (\frac{3}{4}+\delta)n$):] This is similar to Case A.
     With high probability, at least $(1 - \beta) n$ nodes will decide on 1. 

\end{description}
Note that, when a checkpoint falls either under Case A or Case E, with high
probability, it will remain in that case. When a checkpoint falls under Case B,
Case C, or Case D, with constant probability, we get either Case A or Case E in
the following checkpoint. Therefore, in $O(\log n)$ rounds, at least
$(1-\beta)n$ nodes will reach agreement with high probability and all other
nodes will remain undecided.

For property Stability, note that if a node has decided on some value 
in checkpoint $t_R$, it continues to flood its decision message. We showed that, with high probability, at least
$(1-\beta)n$ nodes will decide on the same bit value. Therefore, it follows by
Lemma~\ref{lem:beta} that agreement will be maintained  ad infinitum among at
least $(1 - \beta) n$ nodes. %
\qed
\end{proof}
}

\onlyLong{\subsection{A 3-phase Algorithm for \sa}} \label{sec:obliviousSA}
We will now describe how we use Algorithm~\ref{alg:bc} as a building block for
solving \sa:
In order to use Algorithm~\ref{alg:bc} to solve \sa, we will need to make a
couple of crucial  adaptations. \onlyShort{The first one is needed to guarantee
the Validity property of \sa, while the second one deals with the fact that only
nodes starting with 1 will initiate \bc. These adaptations are fully described in
the attached full paper.}
\onlyLong{
\begin{itemize}
\item Suppose every vertex in $V^0$ has some auxiliary information. We can
  easily adapt Algorithm~\ref{alg:bc} so that when a node $u$ decides on a bit
  value $b$, then, it also inherits the auxiliary information of some $v \in
  V^0$ whose initial bit value was $b$. This is guaranteed because our
  algorithm ensures Validity.
  The auxiliary information can be piggybacked on the messages that $v$ generates throughout the course of the algorithm.

\item For a typical agreement algorithm, we assume that all nodes simultaneously
  start running the algorithm. We want to adapt our algorithm
  so that only nodes in $V^0$ that have an initial output bit of 1 initiate
  the algorithm, while nodes that start with 0 are considered passive, i.e.,
  these nodes do not generate messages themselves, but still forward flooding
  messages and start generating messages from the next checkpoint onward as soon
  as they notice that an instance of the algorithm is running. 
  
  We now sketch how the algorithm can be adapted: In the first checkpoint $t_1$, each node $v$ with a
  1 initiates support estimation and flooding of message $\msg{r_v, b_v=1}$. If
  the number of nodes with 1 is small at checkpoint $t_1$, then, at checkpoint
  $t_2$, nodes that receive estimate values will conclude 0, which will get
  reinforced in subsequent checkpoints.  However, if the number of nodes with a
  1 at checkpoint $t_1$ is large (in particular, larger than $\beta n$), then,
  by suitable flooding, at least $(1-\beta)n$ nodes 
  will know that a support estimation is underway and will participate from
  checkpoint $t_2$ onward.
\end{itemize}
}

\paragraph{Selection and Flooding Phase:} 
  In the very first round, each node $u \in V^0$ generates
  a uniform random number $r_u$ from $(0,1)$ and, if the random number is less
  than $\frac{4\log n}{n}$, $u$ becomes a \emph{candidate} and initiates a message $m_u$ for flooding.
  The
  message $m_u$ contains the random number $r_u$ and the general value $\val_u$ (from domain $\{0,\dots,m\}$) assigned to $u$ by the adversary.
  This phase ends after $\Theta(\log n)$ rounds to ensure that no more than $\beta n$
  nodes are suppressed (the precise bound on the number of rounds is given by  Corollary~\ref{cor:suppress}).
  The flooding of the generated messages, however, goes on ad infinitum.

\paragraph{Candidate Elimination Phase:}
  We  initiate $\Theta(\log n)$ parallel
  iterations of \bc, whereby each iteration is associated with one of the  $\Theta(\log n)$ flooding messages, generated by the candidates in the first phase.
  More precisely, the $i$-th instance of \bc\ for the $i$-th candidate and its flooding message $m_{u_i}$ is designed as follows:
  nodes that have received a flooded message $m_{u_i}$, set
  their input bit (of the $i$-th instance of \bc) to $1$ and initiate \bc.
  We say that a flooded message $m_u$ is
  a {\em survived candidate} message if the instance of \bc\ associated with it reached a
  decision value of $1$.

\paragraph{Confirmation Phase:} 
  Among the survived candidate messages, every node $v$
  chooses the message $m_{u_j}$ among its received messages that has the smallest random number $r_{u_j}$ (and associated general input value
  $\val_{u_j}$), and initiates a support estimation for the number of nodes that have received $m_{u_j}$.
  If the support estimation reveals a support of at least $(1-\beta - \delta)n$ for $m_j$ then $v$ decides on $\val_{u_j}$. 
  Nodes keep flooding their decision ad infinitum.

\begin{theorem}\label{thm:sa}
Consider the oblivious adversary and suppose that $\eps n$
nodes can be subject to churn in every round. The 3-phase
algorithm is correct with high probability and reaches \sa\ in $O(\log^2 n)$
rounds.
\end{theorem}
\begin{proof}
  Validity follows immediately from the fact that nodes only decide on some value that was the input value of a (survived) candidate.

  We now argue Almost Everywhere Agreement:
  Since all nodes choose independently whether to become candidate, a simple application of a standard Chernoff bound shows that the number of candidates is in the range $[2\log n,8\log n]$ with probability $\ge 1-n^{-3}$; in the remainder of this proof, we condition on this event to be true.

  Consider the message $m_u$ generated by some candidate $u$ in the Selection and Flooding phase, and consider its associated instance of \bc:
If $m_u$ has reached at least $(1-\beta)n$ nodes by flooding, it follows by the properties of the \bc\ algorithm that the decision value of \bc\ will be $1$ with probability $1-n^{-2}$.
On the other hand, if $m_u$ has a very small support (say, $\beta n$), the consensus value will be
$0$ with probability $1-n^{-3}$ (cf.\ Case A of the proof of Theorem~\ref{thm:ssd}),
  and, if the support of $m_u$ is neither too small nor too large, the nodes
  will reach consensus on either $0$ or $1$. 
  Thus we can interpret a decision of $1$ regarding the $i$-th message, as a confirmation that the $i$-th candidate had sufficiently large support.
  By taking a union bound, it follows that, with probability at least $1-n^{-2}$, at least $(1-\beta) n$ nodes agree
  on the set of survived candidate messages, since they reached agreement in each iteration of \bc.
Since the adversary is oblivious, each of
the $\Theta(\log n)$ flooding messages generated by the candidates will not be suppressed with probability at least $(1 - \beta)$ (cf.\ Corollary~\ref{cor:random}).
Therefore, with probability $\ge 1 - n^{-2}$, at least one candidate $u$ will have $|\infl(u, O(\log n))| \ge (1 - \beta) n$ and thus the set of survived candidates $S$ will be nonempty; let $w \in S$ be the candidate who generated the smallest random number.
When the support estimation is initiated in the third phase, a set of at least
$(1-\beta)n$ nodes will measure $w$'s support to be at least $(1-\beta - \delta)n$
for some $\delta > 2\beta$ with probability $\ge 1 - n^{-2}$ (cf.\
Theorem~\ref{thm:ObliviousSupport}) and decide on the value $\val_w$ of $w$,
whereas nodes that do not observe high support remain undecided.
This shows Almost Everywhere Agreement. 

Analogously to Algorithm~\ref{alg:bc}, nodes in $S$ flood their decision
messages, which are adopted by newly incoming nodes. By virtue of
Lemma~\ref{lem:beta}, the stability property is maintained ad infinitum.

The additional running time overhead of the above three phases excluding 
Algorithm~\ref{alg:bc} is only in $O(\log n)$. This completes the proof of the Theorem.
\qed
\end{proof}

\section{\sa\ Under an Adaptive Adversary} \label{sec:adaptive}

In this section we consider the \sa\ problem while dealing with a more
powerful adaptive adversary. At the beginning of a round $r$, this adversary
observes the entire state of the network and previous communication between
nodes (including even previous outcomes of random choices!), and thus can adapt
its choice of $G^{r}$ to make it much more difficult for nodes to achieve
agreement. 

It is instructive to consider the algorithms  presented in
Section~\ref{sec:oblivious} in this context.  Both approaches are doomed to fail
in the presence of an adaptive adversary: For the \sa\ algorithm, the expected
number of nodes that initiate flooding in the flooding phase is $\log n$. Even
though each of these nodes would have expanded its influence set to
some constant size by the end of the next round, the adaptive adversary can spot
and immediately churn out all these nodes before they can communicate with
anyone else, thus none of these values will gain any support. 

Algorithm~\ref{alg:bc} fails for the simple reason that the adversary can
selectively suppress the flooding of the smallest generated random value
$z\in\set{1,\dots,n^k}$ with attached bit $b_z$ from ever reaching some 50\% of the
nodes, which instead might use a distinct minimum value $z'$ (with an attached
bit value $b_{z'}\ne b_z$) to guide their output
changes.  

To counter the difficulties we have mentioned, we relax the model. Firstly, we
limit the order of the churn to $\sqrt{n}$. Secondly, we  allow messages of up
to a polynomial (in $n$) number of bits to be sent over a link in a single round. 
Under these relaxations, we can estimate the support of red nodes in the network
simply by flooding all the unique identifiers of the red and non-red
nodes\onlyLong{ (cf.\ Theorem~\ref{thm:AdaptiveSupport})}.

Similarly to Section~\ref{sec:oblivious}, we will first solve \bc\ under these
assumptions and then show how to implement \sa. In this section we assume that
the number of nodes in the network is sufficiently large, such that
\begin{equation} \label{eq:numberOfNodesAdaptive}
  n \gg 4\beta^2\text{.}
\end{equation}
Moreover, we assume that every node can send and process up to $O(n^c + \log m)$
bits per round, where $c$ is a constant and $m$ is the size of the input domain.
\subsection{\bc} \label{sec:adaptiveBC}

\onlyLong{
\begin{algorithm}[h]
  \begin{algorithmic}[1]
  \footnotesize
  \EMPTY Let $decision_u$ be initialized to $\bot$.

  Let $b_u$ be the current output bit of $u$. If $u \in V^0$, then $b_u$ is
  initialized to the input value of $u$; otherwise it is set to $\bot$. 
  
  Let $t_1 = 1$ be the first checkpoint round. Subsequent checkpoint rounds are
  given by  $t_i = t_{i-1} + O(\log^2 n)$, for $i > 1$, with time between
  consecutive checkpoint rounds sufficient for unsuppressed nodes to reach a
  common influence\onlyLong{ (cf. Lemma~\ref{lem:UniversalInfluence2})}. 
  For the terminating checkpoint $t_R$, we choose an $R \in O(\log n)$, i.e.,
  $t_R \in O(\log^3 n)$.
  \EMPTY
  \item[\bf At every checkpoint round $t_i$ \underline{excluding} $t_R$:] 
  \STATE Initiate support estimation (to be completed in checkpoint round $t_{i+1}$).
  \EMPTY 
  \item[\bf At every checkpoint round $t_i$ \underline{excluding} $t_1$, $t_R$:] 
  \STATE Use the support estimation initiated at checkpoint round $t_{i-1}$. Let $\#(1)$ be the estimated support value for nodes that output 1. 
  \IF{support estimation is not accurate within $[\R - \frac{\beta \sqrt{n}}{2}, \R+ \frac{\beta \sqrt{n}}{2}]$}
    \STATE Do nothing. \label{line:doNothing}
  \ELSIF{$\#(1) <  \frac{n}{2}-\frac{\beta \sqrt{n}}{2}$}
    \STATE $b_u \leftarrow 0$.
  \ELSIF{$\#(1) > \frac{n}{2}+\frac{\beta \sqrt{n}}{2}$} \label{line:largeSupport}
    \STATE $b_u \leftarrow 1$.
  \ELSE
    \IF{the outcome of an unbiased coin flip is heads}
      \STATE $b_u \la 0$.
    \ELSE
      \STATE $b_u \la 1$.
    \ENDIF
  \ENDIF
  \EMPTY
  \item[\bf At terminating checkpoint round $t_R$:]
  \IF{$\#(1) \ge  \frac{n}{2}$} \label{line:oneHigh}
   		\STATE $decision_u \la 1$. 
      \STATE Flood a 1-decision message ad infinitum.
  \ELSIF{$\#(0) \ge \frac{n}{2}$}
   		\STATE $decision_u \la 0$. 
   		\STATE Flood a 0-decision message ad infinitum.
  \ENDIF
  \EMPTY
  \item[\bf If $u$ receives a $b$-decision message:]
    \STATE $decision_u\la b$
  \end{algorithmic}
  \caption{\bc\ under an adaptive adversary; code executed by node $u$.}
  \label{alg:adaptive}
\end{algorithm}
\normalsize
}

We now\onlyShort{ briefly} describe an algorithm for solving \bc, which is
similar in spirit to Algorithm~\ref{alg:bc}. \onlyShort{The full statement of
the algorithm can be found in the attached full paper.} The main difference is
the handling of the case where the support of the nodes that output 1 is roughly
equal to the support of the nodes with output bit 0. In this case we rely on the
variance of random choices made by individual nodes to sway the balance of the
support towards one of the two sides with constant probability. 

First, we argue why this technique does not work when the churn limit is $\omega(\sqrt{n})$:
In our algorithm we handle the case where the support of $0$ and $1$ is roughly equal, by causing each node to update its current output bit to the outcome of a (private) unbiased coin flip.
The standard deviation that we get for the sum of these individual random variables is $O(\sqrt{n})$ and the event where the balance is swayed by $O(\sqrt{n})$ occurs with constant probability.
But since the adversary is adaptive and has $\omega(\sqrt{n})$ churn to play with, it can immediately undo this favourable imbalance by churning out nodes such that the support of $0$ and $1$ will yet again be roughly equal. 

\begin{theorem} \label{thm:adaptiveBC}
\onlyLong{Algorithm~\ref{alg:adaptive}}\onlyShort{There is an algorithm that}
solves \bc\ with high probability and reaches stability within $O(\log^3 n)$ rounds, in the presence of
an adaptive adversary and up to $\eps \sqrt{n}$ churn per round.
\end{theorem}
\onlyLong{
\begin{proof}
First consider the Validity property: Suppose that all nodes start with input value~1. 
Theorem~\ref{thm:AdaptiveSupport} guarantees that any node $u$ that receives
insufficiently many identifiers for support estimation, will execute
Line~\ref{line:doNothing} and therefore never set its output to 0. On the other
hand, if $u$ does receive sufficiently many samples, again
Theorem~\ref{thm:AdaptiveSupport} ensures that it will always pass the if-check
in Line~\ref{line:largeSupport}. Thus, no node can ever output 0. The case where
all nodes start with 0 can be argued analogously.

Next, we will show that Algorithm~\ref{alg:adaptive} satisfies Almost
Everywhere Agreement.
Let $N_{i}$ be the number of vertices at checkpoint round $t_{i}$ with output bit 1.
Let $\lo_i$, $\hi_i$, and $\mid_i$, respectively, be the sets of nodes in
$V^{t_i}$ for which $\#(1) \leq  n/2 - \frac{\beta \sqrt{n}}{2}$, $\#(1) \geq  n/2 + \frac{\beta \sqrt{n}}{2}$, and 
$n/2 - \frac{\beta \sqrt{n}}{2} < \#(1) < n/2 + \frac{\beta \sqrt{n}}{2}$; note that nodes are placed in $\lo_i$,
$\hi_i$, and $\mid_i$ based on their $\#(1)$ values, which are estimates of
$N_{i-1}$, not $N_i$. In a slight abuse of notation, we use $\lo_i$, $\mid_i$,
and $\hi_i$ to also refer to their respective cardinalities. Clearly, we have
that 
\begin{equation*}\lo_i + \mid_i + \hi_i = n.\end{equation*}
Furthermore, observe that either $\lo_i$ or $\hi_i$ will be 0. Otherwise, we will have two nodes such that one estimates $N_{i-1}$ below $n/2 - \frac{\beta \sqrt{n}}{2}$, while the other estimates it above $n/2 + \frac{\beta \sqrt{n}}{2}$ --- a violation of Theorem~\ref{thm:AdaptiveSupport}.

Consider the following cases:
\begin{description}
	\item[Case A ($N_{i-1} < n/2 - \beta \sqrt{n}$):] From
    Theorem~\ref{thm:AdaptiveSupport}, $\lo_i \ge n - \beta \sqrt{n}$ and all
    nodes in $\lo_i$ will set themselves to output 0. Once this case is reached
    in some checkpoint, it will be reached in all future checkpoints until $t_R$
    with high probability.
    Therefore, the algorithm guarantees Almost Everywhere Agreement on 0 in
    $t_R$; with high probability, nodes do not pass Line~\ref{line:oneHigh}
    in checkpoint $t_R$, thus no node will ever decide on $1$.
	\item[Case B  ($N_{i-1} > n/2 + \beta \sqrt{n}$):] This case is similar to
    Case A with the difference that almost all nodes decide on 1.
	\item[Case C ($n/2 - \beta \sqrt{n} \le N_{i-1} \le n/2$):] Notice that $\hi_i = 0$. Therefore, 
\begin{equation}
\lo_i + \mid_i \ge n - \beta \sqrt{n}.
\label{eq:LoMid}
\end{equation}
		We consider two subcases: 
  \paragraph{1.} In this case, we assume that $\lo_i$ is at least $n/2 + \beta \sqrt{n}$. This will set $N_i < n/2 - \beta \sqrt{n}$ putting the network in Case A in the next checkpoint.
  \paragraph{2.} In this case, we assume that $\lo_i < n/2 + \beta \sqrt{n}$. This implies that 
 \begin{equation*}\mid_i \ge n - \lo_i - \beta \sqrt{n} \ge n/2 - 2\beta \sqrt{n}.\end{equation*} The nodes in $\mid_i$ will choose 1 or 0 with equal probability. The number of nodes that choose 0 is a binomial distribution with mean $\frac{\mid_i}{2}$ and standard deviation $ \frac{\sqrt{\mid_i}}{2}$. Clearly, with some constant probability, $\frac{\mid_i}{2} + \frac{\sqrt{\mid_i}}{2}$ or more nodes in the set $\mid_i$ will set themselves to output 0. Therefore, with constant probability,  
\begin{align*}
	N_i <  n - \lo_i &- \frac{\mid_i}{2} - \frac{\sqrt{\mid_i}}{2} \\
      < n - \lo_i &- \frac{n - \lo_i - \beta \sqrt{n}}{2} \\
                  &- \frac{\sqrt{n - \lo_i - \beta \sqrt{n}}}{2} 
\end{align*}
Clearly, $N_i < \frac{n}{2} - \beta \sqrt{n}$ if 
\begin{align*}
3 \beta \sqrt{n} &< \sqrt{n - \lo_i - \beta \sqrt{n}},\\
\intertext{which means that}
\lo_i + \beta \sqrt{n} &< n - 9 \beta^2 n.
\end{align*}
We know that $\lo_i < \frac{n}{2} + \beta \sqrt{n}$. Therefore, $N_i < \frac{n}{2} - \beta \sqrt{n}$ if 
\begin{align*}
\frac{n}{2} + 2 \beta \sqrt{n} &< n - 9 \beta^2 n, \\
\intertext{that is,}
2 \beta &< \sqrt{n} \left(\frac{1}{2} - 9 \beta^2 \right ).
\end{align*}
In other words, as long as 
\begin{equation}
n > \frac{4 \beta^2}{\left(\frac{1}{2} - 9 \beta^2 \right )^2},
\label{eq:condition}
\end{equation}
it holds with constant probability that
\begin{equation*}N_i< \frac{n}{2} - \beta \sqrt{n},\end{equation*}
which will put
the network in Case A at the next checkpoint round. Assumption~\eqref{eq:numberOfNodesAdaptive} guarantees that Condition~\eqref{eq:condition}
is easily met. 
\item[Case D ($n/2  < N_{i-1} \le n/2 +\beta \sqrt{n}$):] Using arguments
similar to Case C, we can show that with constant probability, 
\begin{equation*}N_i > \frac{n}{2} + \beta \sqrt{n},\end{equation*}
thereby, putting the network in Case B. 
\end{description}
Clearly, after $O(\log n)$ checkpoint rounds the network
will reach either Case A or Case B\footnote{Due to Equation
\eqref{eq:numberOfNodesAdaptive} we know that Cases A and B exist.} with high probability and hence achieve Almost Everywhere
Agreement on either 0 or 1.

For property Stability, note that if a node has decided on some value $\ne \bot$
in checkpoint $t_R$, it continues to flood its decision message. Since at least
$(1-\beta)n$ have decided, it follows by Lemma~\ref{lem:beta} that any nodes
that have been churned in will also decide on this value within a constant
number of rounds, thus agreement will be maintained ad infinitum.
\qed
\end{proof}
}

\onlyLong{\subsection{\sa}} \label{sec:adaptiveSA}
Now that we have a solution for \bc, we will show how to use it to solve \sa\
where nodes have input values from some set $\set{0,\dots,m}$, for $m\ge 1$.
Given some input value $\val$ we can write it in the base-2 number system as
$(b_0,\dots,b_{\log m})$ where $b_i\in\set{0,1}$, for $0\le i\le \log m$. We
call $\val$ a \emph{general input value} and $b_i$ a \emph{binary input value}.

The basic idea of the \sa\ algorithm is to run an instance of the \bc\ algorithm
for each $b_i$ and then combine the agreed bits $d_1,\dots,d_{\log m}$ to obtain agreement on the
general input values. We now describe our algorithm; the detailed pseudo code is
presented in Algorithm~\ref{alg:sa}.
Consider the $i$-th iteration of Algorithm~\ref{alg:sa} and suppose that $d_1,\dots,d_{i-1}$ are the first $i-1$ decision values of the previous $i-1$ iterations of the \bc\ algorithm.
We say that a node $u$ \emph{knows a general input value matching the first $i$ binary decision values}, if $u$ has knowledge of a some $\val \in \{0,\dots,m\}$ that was the input value of some node $v$ and the first $i-1$ bits of $\val$ are exactly $d_1,\dots,d_{i-1}$.
We denote the $i$-th bit value of a general value $\val$ by $\val[i]$.
Recall that the \bc\ algorithm executes the support estimation routines developed in Section~\ref{sec:supportAdaptive}.
We slightly modify the support estimation routine by requiring each node $u$ to also piggyback its current general value $\val_u$ onto the message it generates for support estimation.
Moreover, when $u$ floods the decision message of the \bc\ algorithm, it also piggybacks $\val_u$.
Whenever a node $v$ updates its current output bit value to $b$, this guarantees that $v$ has learned of a general value $\val_w$ that has $b$ as its first bit.
Thus $v$ sets $\val_v$ to the new value $\val_w$ and chooses its next input value for the $(i+1)$-th iteration of the \bc\ algorithm to be the $(i+1)$-th bit of $\val_v$.
This is formalized in the following lemma:
\begin{lemma} \label{lem:validity}
  Consider iteration $i$ of the \bc\ subroutine executed in Algorithm~\ref{alg:sa}.
  If a node $u$ has a current binary output value of $b$, then the $i$-th bit of $\val_u$ is $b$.
\end{lemma}
\begin{proof}
  We will show the result by induction over the iterations of the \bc\ algorithm. 
  Initially, in the first iteration, node $u$ uses the first bit of its input value $\val_u$.
  Now suppose that $u$ sets its output bit to $1-b$ at some point during the first iteration.
  We say that $u$ violates \emph{general validity}.
  There are two possible cases: In the first case, $u$ observed a sufficiently large support for $1-b$ and thus received a support estimation message generated by a node $v$ that had a current output bit $1-b$, while in the second case $u$ received a decision message generated by $v$.
  In either case, it follows from the description of the algorithm that node $v$ has piggybacked $\val_v$ on top of this message.
  If $\val_v[i]=1-b$, then $v$ has updated its own output bit without updating $\val_v$, due to receiving some message from another node $v'$, and both nodes, $v$ and $v'$, violate general validity.
 By backwards traversing this chain of nodes that violate general validity, we eventually reach a node $w$ which has set its output bit value to $1-b$ but $\val_w[i]=b$, without having received a message from a node that violates general validity.
  According to the \bc\ algorithm, $w$ only sets its bit value to $1-b$ if it has either observed sufficient support for $1-b$ or received a decision message containing a value of $1-b$.
  In both cases, it follows from the description of the algorithm that $w$ updates $\val_w$ to the piggybacked general value, the $i$-th bit of which is $1-b$, providing a contradiction.
  \qed
\end{proof}
The above lemma guarantees that we can combine the decision bits of the \bc\ iterations to get a general decision value that satisfies validity. 
We can therefore show the following theorem:
\onlyLong{
\begin{algorithm}[h]
  \begin{algorithmic}[1]
  \footnotesize

	\STATE Suppose that node $u$ starts with an initial general input value $\val_u$.

	\FOR{$i \leftarrow 0 \text{ to } \log m$}
  \STATE Node $u$ initiates \bc\ by proposing the $i$-th bit of its current $\val_u$. Recall that \bc\ will be reached in $O(\log^3 n)$ rounds.
		\STATE When participating in the support estimation that is part of \bc\, each node $u$ piggybacks its $\val_u$ on top of the support estimation message.

    \STATE Let $d_i$ be the decision returned by \bc\ algorithm.
    If node $u$ has decided on bit value $d_i \in \{0,1\}$, then $u$ has learned of a general input value $\val$ where the $i$-th bit is $d_i$:
    Node $u$ updates its current value $\val_u$ by setting it to $\val$ and floods $\val_u$ along the decision message according to the \bc\ algorithm.
    \STATE If $u$ did not decide in the \bc\ algorithm, then $u$  does not propose a value in the $(i+1)$-th iteration.
  \ENDFOR
  \STATE If $u$ did not decide in the last iteration, it remains undecided.
  Otherwise, $u$ returns the $\val_u$ as its decision value and floods this value ad infinitum.
  \end{algorithmic}
  \caption{Solving \sa\ using \bc. Pseudo code for node $u$.}
  \label{alg:sa}
\end{algorithm}
\normalsize
}

\begin{theorem} \label{thm:adaptiveSA}
  Suppose that the network is controlled by an adaptive adversary who can
  subject up to $\eps\sqrt{n}$ nodes to churn in every round.
  There is an algorithm that solves \sa\ with high probability and reaches stability in $O(\log m \log^3 n)$.
\end{theorem}
\begin{proof}
  Almost-everywhere agreement follows almost immediately from the fact that the \bc\ algorithm satisfies almost-everywhere agreement; what remains to be shown is that all except $\beta \sqrt{n}$ nodes decide:
  Note that it is possible that a set $S$ of $\beta \sqrt{n}$ nodes can remain undecided when running an instance of the \bc\ algorithm.
  The nodes in $S$ will not propose any values in the next iteration but will participate in the support estimation and the propagation of messages.
  By the correctness of the \bc\ algorithm, all except $\beta \sqrt{n}$ nodes eventually know the decision bit $d_i$ of the $i$-th iteration.
  In the next iteration, any node $v$ that knows the decision bit $d_{i+1}$, also knows a general value $\val$ such that $\val[i+1]=d_{i+1}$ and thus can propose in the subsequent iteration.
  This holds regardless of whether $v \in S$ and thus all except $\beta\sqrt{n}$ nodes participate in each iteration.

  For validity, we argue that Algorithm~\ref{alg:sa} maintains the following invariant at the end of every iteration $i$: a node that is aware of the decision (bit) values of the first $i$ runs of the \bc\ subroutine, has knowledge of a general value matching the first $i$ binary decision values.
  By Lemma~\ref{lem:validity}, it follows that if a node $u$ proposes a bit $b$ in iteration $i$, then $b$ is the $i$-th bit of some general input value $\val$.
  This guarantees that the sequence of decision bits correspond to some general input value and thus satisfies validity.

  Finally, we observe that the proof of stability is identical to the \bc\ algorithm, thus completing the proof.
  \qed
\end{proof}

\section{Impossibility of a Deterministic Solution} \label{sec:impossibility}

In this section we show that there is no deterministic algorithm to solve \sa\ even
when the churn is restricted to only a constant number of nodes per round. As a consequence, randomization is a necessity for solving \sa.  
\onlyLong{

We introduce some well known standard notations (see \cite[Chap.\ 5]{AW04})
used for showing impossibility results of agreement problems.
The \emph{configuration} $C^r$ of the network at round $r$ consists of 
\begin{compactitem}
  \item the graph of the network at that point in time, and 
  \item the local state of each node in the network. 
\end{compactitem}
A specific run $\rho$ of some \sa\ algorithm $\A$ is entirely
determined by an infinite sequence of configurations $C^0,C^1,\dots$ where $C^0$
contains the initial state of the graph before the first round.
Consider the input value domain $\set{0,1}$.
A configuration $C^r$ is \emph{1-valent} (resp.,
\emph{0-valent}) if all possible runs of $\A$ that share the common prefix
up to and including $C^r$, lead to an agreement value of 1 (resp., 0). Note
that this decision value refers to the decision of the large majority of
nodes; strictly speaking, a small fraction of nodes might remain undecided on $\bot$.
A configuration is
\emph{univalent} if it is either 1-valent or 0-valent.  Any configuration that
is not univalent is called a \emph{bivalent} configuration.  
The following observation follows immediately from the definition of the \sa\ problem.

\begin{observation} \label{lem:bivalent} Consider a bivalent configuration $C^r$ in round $r$ reached by an algorithm $\A$ that solves \sa\ and ensures Almost Everywhere Agreement.
No node in $V^r$ can have decided on a value $\ne\bot$ by round $r$.
\end{observation}
}
\begin{theorem}\label{thm:impossibility}
Suppose that the sequence of graphs $(G^r)_{r\ge 0}$ is an expander family with
maximum degree $\Delta$. Assume that the churn is limited to at most $\Delta$+1
nodes per round. There is no deterministic algorithm that solves \sa\ if the
network is controlled by an adaptive adversary.
\end{theorem}
\onlyLong{
\begin{proof}
We use an argument that is similar to the argument used in the proof that
$f+1$ rounds are required for consensus in the presence of $f$ faults (cf.\
\cite[Chap.\ 5]{AW04}).
For the purpose of this impossibility proof, we restrict the input domain of
nodes to $\set{0,1}$ and allow arbitrary congestion on the communication
channnels.  Moreover, we assume that the topology of the network is fixed
throughout the run. Thus the adversary can only ``replace'' nodes at the same
position by some other nodes.

For the sake of contradiction, assume that such a deterministic algorithm $\A$
exists that solves \sa\ under the assumed settings. 
We will prove our theorem by inductively constructing an
infinite run $\rho$ of this algorithm consisting of a sequence of bivalent
configurations. By virtue of Observation~\ref{lem:bivalent} this allows us to conclude
that nodes do not reach almost everywhere agreement.

To establish the basis of our induction, we need to show that there is an
initial bivalent configuration $C^0$ at the start of round 1. Assume
in contradiction that there is no bivalent starting configuration.
Let $D_0$ (resp.\ $D_1$) be the configuration where all nodes
start with a value 0 (resp., 1); note that by validity the decision 
value must be on 0 (resp., 1).
Consider the sequence of configurations starting at $D_0$ and ending at $D_1$ where the only difference between any two configurations adjacent (in this sequence) is a single bit, i.e., exactly $1$ node has a different input value.
Since $D_0$ is $0$-valent and $D_1$ is $1$-valent, this implies that there are two possible
starting configurations in this sequence, $C^0_0$ and $C^0_1$, in which (i) the input values are
the same for all but one node $u^0$, but (ii) $C^0_0$ is 0-valent whereas
$C^0_1$ is 1-valent. Consider the respective one-round extension of $C^0_0$ and
$C^0_1$ where the adversary simply churns out node $u^0$.
Both successor configurations $C^1_0$ and $C^1_1$ are indistinguishible for all
other nodes, in particular they have no way of knowing what initial value was
assigned to $u^0$, since all witnesses have been removed by the adversary.
Therefore, $C^{1}_0$ and $C^1_1$ must both be either 0-valent or 1-valent, a
contradiction. This shows that there is an initial bivalent configuration,
thereby establishing the basis for our induction. 

For the inductive step, we assume that the network is in a bivalent
configuration $C^{r-1}$ at the end of round $r-1$. We will extend $C^{r-1}$ by
one round (guided by the adversary) that yields another bivalent configuration
$C^{r}$. Assume for the sake of a contradiction that every possible one-round
extension of $C^{r-1}$ yields a univalent configuration.  Without loss of
generality, assume that the one-round extension $\gamma$ where no node is
churned out is 1-valent and yields configuration $C^{r}_1$.  Since by assumption
$C^{r-1}$ was bivalent, there is another one-round extension $\gamma'$ that
yields a 0-valent configuration $C^{r}_0$. Moreover,  we know that a nonempty
set $S$ of size at most $\Delta$+1 nodes must have been subject to churn in
$\gamma'$. (This is the only difference between $C^{r}_0$ and $C^r_1$ ---
recall that the edges of the graph are stable throughout the run.)

Let $S'$ be a subset of $S$ and let
$\gamma_{S'}$ be the one-round extension of $C^{r-1}$ that we
get when only nodes in $S'$ are churned out. Clearly,
$\gamma=\gamma_{\emptyset}$ and $\gamma'=\gamma_S$.
Consider the lattice of all such one-round extension bounded by $\gamma$ and
$\gamma'$ that is given by the power set of $S$. Starting at $\gamma$ and moving
towards $\gamma'$ along some path, we must reach a one-round extension
$\gamma_{\set{v_1,\dots,v_k}}$ that yields a 1-valent configuration
$D^{r}_1$, whereas the next point on this path is some one-round extension
$\gamma_{\set{v_1,\dots,v_{k+1}}}$ that ends in a 0-valent configuration
$D^{r}_0$. The only difference between these two extensions is that node
$v_{k+1}$ is churned out in the latter but not in the former extension. Now
consider the one-round extensions of $D^r_0$ and $D^r_1$ where $v_{k+1}$ and all
its neighbors are churned out, yielding $D^{r+1}_0$ and $D^{r+1}_1$. For all
other nodes, $D^r_0$ and $D^{r}_1$ are indistinguishible and therefore they must
either both be 0-valent or both be 1-valent. This, however, is a contradiction.
\qed
\end{proof}
}
\onlyLong{
Considering that expander graphs usually are assumed to have constant degree,
Theorem~\ref{thm:impossibility} implies that even if we limit the churn to a
constant, the adaptive adversary can still beat any deterministic algorithm.
}

\section{Conclusion}
We have introduced a novel framework for analyzing highly dynamic
distributed systems with churn. We believe that our model
captures the core characteristics of such systems: a large amount of churn
per round and a constantly changing network topology.  Future work  involves extending our model to include Byzantine nodes and corrupted
communication channels. Furthermore, our work raises some key questions:
How much churn can we tolerate in an adaptive setting? Are there algorithms that
tolerate linear (in $n$) churn in an adaptive setting? We show that we can tolerate $O(\sqrt{n})$ 
churn in an adaptive setting, but it takes a polynomial (in $n$) number of communication bits per round.
An intriguing problem is to reduce the number of bits to polylogarithmic in $n$.

While the main focus of this paper was achieving agreement among nodes
which is one of the most important tasks in a distributed system, as a next step, it might be worthwhile to investigate whether the techniques presented in this paper can serve  as useful building blocks for tackling other important tasks
like aggregation or leader election in highly dynamic networks.

\bibliographystyle{plain}
\bibliography{papers}

\end{document}